\documentclass{aa}

\usepackage{graphicx}
\usepackage{ulem}
\usepackage{cancel}
\usepackage{hyperref}
\usepackage{txfonts}
\usepackage[utf8]{inputenc}
\usepackage{amsmath}	
\usepackage{amssymb}	
\usepackage{subcaption}
\usepackage{gensymb}
\usepackage[flushleft]{threeparttable}
\usepackage[pdftex,usenames,dvipsnames]{xcolor}
\usepackage{natbib}
\usepackage{aas_macros}
\usepackage{url}
\usepackage[none]{hyphenat}
\usepackage{times}
\usepackage{array}
\usepackage{lscape}
\usepackage[colorinlistoftodos]{todonotes}
\usepackage{rotating}
\usepackage{adjustbox} 

\begin{document}

   \title{MaNGA AGN dwarf galaxies (MAD)\\ II. AGN outflows in dwarf galaxies}

   \author{V. Rodr\'{i}guez Morales,\thanks{E-mail: victorproyecto98@gmail.com}\inst{1} 
          M. Mezcua,\inst{1,2}
          H. Dom\'{i}nguez S\'{a}nchez,\inst{3,4}
          A. Audibert, \inst{5,6}
          F. M\"{u}ller-S\'{a}nchez, \inst{7}
          M. Siudek, \inst{1,5}
          \and
          A. Er\'{o}stegui \inst{1}
          }

   \institute{\textsuperscript{1} Institute of Space Sciences (ICE, CSIC), Campus UAB, Carrer de Magrans, 08193 Barcelona, Spain\\
   \textsuperscript{2} Institut d'Estudis Espacials de Catalunya (IEEC), Edifici RDIT, Campus UPC, 08860 Castelldefels (Barcelona), Spain \\
   \textsuperscript{3} Centro de Estudios de F\'{i}sica del Cosmos de Arag\'{o}n (CEFCA), Plaza San Juan, 1, 44001, Teruel, Spain \\
   \textsuperscript{4} Instituto de F\'{i}sica de Cantabria, Avenida de los Castros, s/n, 39005 Santander, Cantabria, Spain \\
   \textsuperscript{5} Instituto de Astrof\'{i}sica de Canarias, Calle V\'{i}a L\'{a}ctea, s/n, E-38205 La Laguna, Tenerife, Spain \\   
  \textsuperscript{6} Departamento de Astrof\'{i}sica, Universidad de La Laguna, E-38206 La Laguna, Tenerife, Spain \\   
   \textsuperscript{7} Department of Physics and Materials Science, The University of Memphis, 3720 Alumni Avenue, Memphis, TN 38152, USA
   }

   \date{Received XX XX, XXXX; accepted XX XX, XXXX} 

 
  \abstract
   {Feedback from an active galactic nucleus (AGN)  is one of the most important mechanisms in galaxy evolution. This phenomenon is usually found in massive galaxies and is known to regulate star formation. Although dwarf galaxies are assumed to be regulated by supernova feedback, recent studies have offered evidence to support the presence of AGN outflows and feedback in dwarf galaxies.}
   { We investigate the presence of AGN outflows in a sample of 2292 dwarf galaxies with AGN signatures drawn from the MaNGA survey. Thanks to the integral field unit data from MaNGA, we are able to spatially resolve these outflows and study their kinematics and energetics.}
   {Using the Galaxy/AGN Emission Line Analysis TOol (GELATO) Python code, we fit the AGN-stacked spectrum of each galaxy. This is the stack of all the spaxels classified as AGNs or composites based on their emission line diagnostic diagrams and, in particular, the [OIII]$\lambda$5007\AA\ emission line. If the galaxies exhibited a broad [OIII] emission line component in the stacked spectrum, we  ran GELATO through all the spaxels classified as AGNs and composites in the emission line diagnostic diagrams.}
   {We found 13 new dwarf galaxies that present outflow signatures based on the presence of a broad [OIII] emission line component. Their velocity measurement W$_{80}$ (width containing 80$\%$ of the flux of the [OIII]$\lambda$5007\AA\ emission line) ranges from 205 to 566 km s$^{-1}$ and the kinetic energy rate ranges from $\sim10^{35}$ to $\sim10^{39}$ erg s$^{-1}$. Stellar processes are unlikely to explain these outflow kinetic energy rates in the case of nine dwarf galaxies. We found a correlation between the W$_{80}$ velocity and the [OIII] luminosity as well as between the kinetic energy rate of the outflow and the bolometric luminosity spanning from massive to dwarf galaxies. This suggests a similar behaviour between the AGN outflows in the dwarf galaxy population and those in massive galaxies.
 }
   {}

    \titlerunning{AGN outflows in dwarf galaxies}
    \authorrunning{Rodr\'{i}guez Morales et al.}
   \maketitle

\section{Introduction}

Although active galactic nuclei (AGNs) are mostly found in massive galaxies, in the late 1980s \cite{filippenko1989discovery} discovered an AGN in the dwarf (stellar mass $M_{\ast}\leq  10^{10}M_{\odot}$) spiral galaxy NGC 4395. Nowadays, the use of optical spectroscopy using surveys such as the Sloan Digital Sky Survey (SDSS) or the Dark Energy Spectroscopic Instrument (DESI) has allowed us to find thousands of dwarf galaxies hosting AGNs (see \citealt{pucha2025tripling}). This has been mostly based on the use of emission line diagnostic diagrams such as the Baldwin, Phillips, and Terlevich (BPT; \citealt{baldwin1981classification}), which enable us to distinguish between gas ionisation produced by AGNs, stars, or both. This distinction can be made by comparing the ratio of the narrow emission lines [OIII]$\lambda$5007/H$\beta$ against [NII]$\lambda$6583/H$\alpha$, [SII]$\lambda\lambda$6716, 6731/H$\alpha,$ or [OI]$\lambda$6300/H$\alpha$ (\citealt{kauffmann2003host}; \citealt{kewley2001theoretical, kewley2006host}). AGNs in massive galaxies often exhibit a broad H$\alpha$ or H$\beta$ component originating from dense gas clouds in the broad-line region (BLR) located a few parsecs from the supermassive black hole (SMBH; $M_\mathrm{BH} > 10^6 M_{\odot}$). These clouds are ionised by continuum emission from the accretion disk of the SMBH. In massive galaxies, the typical full width at half maximum (FWHM) of the broad component ranges from $\sim$1000 to $\sim$5000 km s$^{-1}$ (\citealt{osterbrock2006astrophysics}; \citealt{kollatschny2013shape}). In dwarf galaxies, such broad components have also been identified, but with a less pronounced width of $\sim$500-1000 km s$^{-1}$ (e.g. \citealt{reines2013dwarf}; \citealt{chilingarian2018population}). Assuming that the gas producing the broad emission line is moving in circular orbits in the BLR around the SMBH, a BH mass measurement can be obtained from the detection of broad Balmer lines. In the case of dwarf galaxies, this BH mass is typically $M_\mathrm{BH}=10^5-10^6M_\odot$; therefore, it is lower than that of SMBHs (e.g. \citealt{greene2007new}; \citealt{reines2013dwarf}; \citealt{moran2014black}; \citealt{baldassare201550}; \citealt{marleau2017infrared}; \citealt{chilingarian2018population}; \citealt{mezcua2020hidden, mezcua2024manga};  \citealt{salehirad2022hundreds}).\\


BPT diagnostic diagrams used in surveys that only obtain the spectrum at the centre of the galaxy can miss AGN in galaxies with a high star formation rate (SFR) that dilutes the AGN signatures or in cases where the AGN is displaced from the centre of the galaxy \citep{comerford2014offset}. This displacement is expected to occur in  $\sim 50\%$ of dwarf galaxies (e.g. \citealt{bellovary2019multimessenger}). This problem can be solved by observing galaxies in the X-ray and radio bands (e.g. \citealt{hasinger2008absorption}; \citealt{mezcua2016population, mezcua2018intermediate, mezcua2019radio};  \citealt{hickox2018obscured}; \citealt{reines2020new}; \citealt{birchall2020x}; \citealt{bykov2024srg}). Alternatively, integral field unit (IFU) spectographs allow us to identify these `hidden AGN' because they provide information spaxel by spaxel (i.e we have a spectrum for each pixel of the galaxy) thanks to the different fibers arranged of each IFU (e.g. \citealt{ricci2014integral}; \citealt{da2017ngc}; \citealt{wylezalek2020ionized}; \citealt{liu2020integral};  \citealt{mezcua2020hidden, mezcua2024manga}). Furthermore, by carrying out spatially-resolved emission line diagnostic diagrams (i.e. BPT spaxel by spaxel), we are able to identify which regions of the galaxy are ionised by AGNs or star formation (SF) processes (e.g.\citealt{mezcua2020hidden, mezcua2024manga}; \citealt{wylezalek2020ionized}).\\  

AGNs release energy in the form of radiation or of mechanical radio plasma coming from the gravitational energy of the material accreted by the BH. There are different launching mechanisms, such as collimated jets of charged particles or winds formed from the radiation pressure or from the accretion disk. These winds rise up to outflows when they interact with and swept up material from the interstellar medium (\citealt{harrison2024observational}). Outflows can also be produced by SF processes (e.g. \citealt{shepherd2007molecular}; \citealt{gatto2017silcc}; \citealt{romano2023star}; \citealt{sau2023star}) and can be identified by looking for broadened or shifted components in emission lines like [OIII]$\lambda$5007, [NII]$\lambda$6583 or [SII]$\lambda\lambda$6716, 6731 (e.g. \citealt{holt2008fast}; \citealt{liu2013observations}; \citealt{harrison2014kiloparsec}; \citealt{leung2019mosdef}; \citealt{wylezalek2020ionized}). By combining the kinematic information of the emission lines with the BPT diagram diagnostics to access the gas excitation, we have access to a powerful way to infer the origin of the outflows and determine whether these are AGN or SF processes. \\

The impact of AGN outflows on the host galaxy, known as AGN feedback, plays a role in regulating the star formation due to the injection of energy and the momentum transfer in the interstellar medium of the galaxy. As a result, the growth of SMBHs is interconnected with the evolution of their host galaxies, a phenomenon referred to as BH-galaxy co-evolution (e.g. \citealt{kormendy2013coevolution}; \citealt{zhuang2023evolutionary}; \citealt{capelo2024black}). AGN feedback and outflows are typically found in massive galaxies and they are generally suggested to prevent gas from cooling and to redistribute it (e.g. \citealt{croton2006many}; \citealt{bower2006breaking}; \citealt{falceta2010precessing}). Nevertheless, high spatial resolution observations and recent simulations have made the understanding of AGN feedback more complex, pointing out the importance of initial conditions in different factors, such as the gas distribution or the coupling between the outflow and the interstellar medium (e.g. \citealt{tanner2022simulations}, \citealt{clavijo2024role}, see \citealt{harrison2024observational} for a recent review).   \\

In the low-mass regime, the presence of AGN outflows is a matter of debate. AGNs are typically fainter and less powerful in dwarf galaxies than in massive galaxies (but see \citealt{mezcua2023overmassive}, \citealt{mezcua2024overmassive}). Furthermore, star formation in dwarf galaxies is commonly assumed to be regulated by stellar winds coming from supernovae (SNe). However, recent studies show evidence for the presence of AGN outflows and feedback in dwarf galaxies. \cite{penny2018sdss} found evidence of AGN feedback in a subset of 69 quenched low-mass galaxies selected from the first two years of the SDSS Mapping Nearby Galaxies at Apache Point Observatory (MaNGA; \citealt{bundy2015overview}) survey. \cite{manzano2019agn} reported  the detection of six dwarf galaxies selected from SDSS that have an outflow component in the [OIII] doublet with emission lines ratios consistent with AGN ionisation.  \cite{liu2020integral} used a sample of eight dwarf galaxies with known AGN taken from \cite{manzano2019agn}, and the IFU data reveals signs of outflowing gas in seven of them based on the detection of a wing in the Gaussian fit of the [OIII] line. \cite{liu2024fast} studied three far-ultraviolet dwarf galaxies. They reported a fast outflow detected in multiple transitions in one source. In another source, they detected a blue-shifted [HeII]$\lambda$1640\AA\ emission line, likely tracing a highly ionised AGN outflow. \cite{zheng2023escaping} detected an escaping outflow in a dwarf galaxy with an intermediate-mass black hole (IMBH, $100M_{\odot}\leq M_{\mathrm{BH}}\leq 10^6 M_{\odot}$) and studied\ the importance of the feedback by comparing the size of the outflow with the extended narrow-line region. \cite{wang2024rubies} reported a low-mass galaxy at redshift z$=$3.1 with AGN signatures and with an absorption feature in the wings of [HeI]$\lambda$10839\AA,\ indicating an ionised gas outflow. \cite{salehirad2025ionized} reported 11 AGN dwarf galaxies with fast outflows consistent with an AGN-driven origin. \\


On the theoretical side, cosmological simulations considering AGN feedback in high-redshift dwarf galaxies indicate that BHs can provide a significantly amount of feedback that can quench SF \citep{barai2018intermediate}. In contrast, \cite{trebitsch2018escape} found that even in the most extreme BH growth scenario, SNe feedback quenches the AGN feedback, with this latter type shown to be negligible in comparison. In recent simulations, \cite{koudmani2022two} found a midpoint where a more moderate SNe feedback in combination with an efficient AGN could be an alternative. In this simulation, a variety of outcomes can be obtained depending on the accretion model: from no additional suppression to moderate regulation of SF to catastrophic quenching. \cite{arjona2024role} investigated  magneto-hydrodynamical simulations of the formation of dwarf galaxies: one with AGN feedback and the other with AGN feedback turned off. The AGN runs reproduce satisfactorily several scaling relations and they find that the global SF of galaxies with AGN is reduced compared to those galaxies in which the AGN has been turned off. At this point, different cosmological simulations bring on different results; thus, observational constraints are needed. Local dwarf galaxies have presumably not experienced any significant evolution and thus resemble those first galaxies formed in the early Universe (z $\sim 12$). Therefore, understanding feedback processes and their contribution in dwarf galaxies could be an essential key for cosmological models of galaxy evolution.\\


In this paper, we present a new sample of 13 dwarf galaxies with AGN outflow candidates drawn from the final MaNGA data release (DR17) \footnote{\url{https://www.sdss4.org/dr17/manga/}}. We compare their properties with those of massive galaxies with AGN-driven outflows. The paper is organised as follows. In Sect. \ref{section:sample}, we study a sample of 2292 dwarf galaxies from \cite{mezcua2024manga} that present AGN ionisation signatures in the spatially-resolved BPTs. We analyse the [OIII]$\lambda$5007\AA\  line in the stacked spectrum, followed by the individual spaxels, to detect and spatially resolve the outflows. In Sect. \ref{section:results}, we derive the extension of the outflows, study their kinematic and energetic properties and discuss their AGN origin. Final conclusions are drawn in Sect. \ref{section:conclu}.  Throughout the paper a standard $\Lambda$CDM cosmology is adopted with $H_0=70$ km s$^{-1}$  Mpc$^{-1}$, $\Omega_\mathrm{M}=0.3$, and $\Omega_{\Lambda}=0.7$. \\



\section{Sample and analysis}
\label{section:sample}
\subsection{Sample of dwarf galaxies in MaNGA}
\label{section:dwarfsample}
The sample studied in this paper is drawn from \cite{mezcua2024manga}. They made a selection of 3306 dwarf galaxies ($M_{\ast}\leq  10^{10}M_{\odot}$) from the MaNGA DR17, which is the final MaNGA data release that includes IFU observations and data products of more than 10000 nearby galaxies. Each IFU is composed of tightly packed arrays of 19 to 127 optical fibres, with diameters from 12 to 32 arcsec. It covers a wavelength range of 3600-10000\AA \  with a spectral resolution of R$\sim$2000. \\

\cite{mezcua2024manga} made a pioneering classification of the spaxels of the 3306 dwarf galaxies based on the [NII]-BPT, [SII]-BPT and [OI]-BPT diagrams using the \cite{kauffmann2003host} classification cutoff, the \cite{kewley2001theoretical} and the \cite{kewley2006host} maximum starburst lines. They combine this classification with the WHAN diagram, used to distinguish between emission by AGN or by hot old stars (\citealt{fernandes2010alternative}). This yields 2292 dwarf galaxies with strong AGN ionisation signatures (i.e. more than 20 spaxels with a signal-to-noise ratio (S/N $\geq 3$) classified as AGN, SF-AGN (i.e. SF in the [NII]-BPT but AGN in the [SII]-BPT or [OI]-BPT), composite, or LINER\footnote{low-ionisation emission line region (LINER)}. For a detailed description, see \citealt{mezcua2024manga}).\\

\begin{figure*}
\centering
\includegraphics[width=\textwidth]{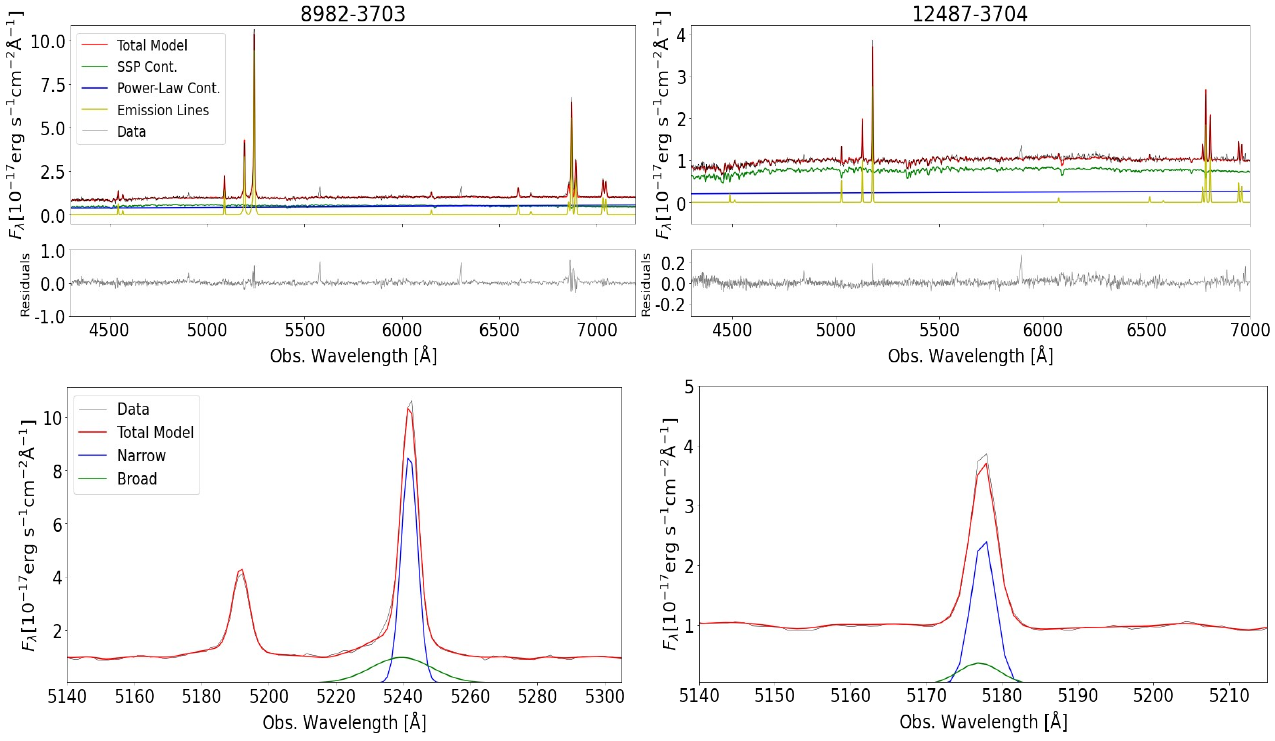}
\caption{Top left and right: Spectra of the galaxies 8982-3703 and 12487-3704, respectively. The dark line is the stacked spectrum of the AGN spaxels. The green line is the best fit of the stellar continuum combining SSPs. The blue line is an additional power-law component, the yellow line represents the fitted emission lines, and the red line is the sum of all these components that compose the total model. The residuals after subtraction of the total model from the data are shown at the bottom. Bottom-left and right: Zoom-in on the [OIII]$\lambda5007$\AA\ doublet region of the stacked spectrum. The [OIII] line is fitted by a narrow (blue) and a broad (green) component.}
\label{fig:out and narr}
\end{figure*}

For all these 2292 dwarf galaxies, \cite{mezcua2024manga} stacked the spectrum of those AGN and composites spaxels. The stacking procedure is described in Appendix A in \cite{mezcua2020hidden}. In this paper, the stacked spectrum of these galaxies is fitted using the Galaxy/AGN Emission Line Analysis TOol (GELATO; \citealt{hviding2022new}), available on Zenodo\footnote{\url{https://zenodo.org/records/5831730}}. GELATO is a Python code designed to retrieve the kinematics and line fluxes of optical spectra. It fits multiple components to lines with complex kinematics and allows the user to tie parameters between emission lines (e.g. H$\alpha$ and H$\beta$ velocity dispersion or the flux ratio between [OIII]$\lambda\lambda5007$, 4959\AA\ emission lines which are known due to the transition probabilities). To run GELATO, the spectral flux, the wavelength, the inverse-variance (ivar), and the redshift (taken  from the NASA-Sloan Atlas (NSA) catalogue\footnote{\url{http://nsatlas.org/data}}) must be provided in addition to the emission lines to be fitted, which are: [SII]$\lambda\lambda6716$, 6731\AA, [NII]$\lambda\lambda6583$, 6548\AA, H$\alpha$$\lambda6563$\AA, [OI]$\lambda\lambda6300$, 6364\AA, HeI$\lambda5876$\AA, [OIII]$\lambda\lambda\lambda5007$, 4959, 4364\AA, H$\beta$$\lambda4861$\AA, [H$\gamma$]$\lambda4340$\AA, [NeIII]$\lambda3869$\AA, and [OII]$\lambda\lambda3726$, 3728\AA. GELATO models the stellar component using the Extended MILES stellar library (\citealt{vazdekis2016uv}) combining simple stellar populations (SSP). An additional power-law component is added if the continuum model passes a statistical F-Test at a 3$\sigma$ level, corresponding to a likelihood of $\sim 99.87\%$ (\citealt{hviding2022new}). Additional components are also tested with an F-Test at a 3$\sigma$ level and the final model is the one with the lowest Akaike information criteria (AIC; \citealt{akaike1974new}), which is an estimator of the relative quality of statistical models. We allow GELATO to fit a broad component to the Balmer lines and an outflow component to the [OIII] emission line, apart from the narrow component (all of them fitted as Gaussians). The distinction between broad and outflow component is based on the bounds of the dispersion and redshifts of these components. For the outflow component, the dispersion has a lower value of 100 km s$^{-1}$ and an upper value of 750 km s$^{-1}$ corresponding to the 1st and 99th percentile values of the dispersion of the second [OIII] component (\citealt{mullaney2013narrow}, for further details see \citealt{hviding2022new}). In this paper we use the terms `broad component' and `outflow component' interchangeably.\\

Once the stacked spectrum of each of the 2292 galaxies was fitted, the [OIII]$\lambda5007$\AA\ emission line was examined to determine whether a broadened or shifted component can be observed (see Fig. \ref{fig:out and narr}), indicating gas with different kinematics than the rest of the ionised gas in the galaxy and thus potentially tracing the outflowing gas (e.g. \citealt{liu2013observations}; \citealt{manzano2019agn}). Furthermore, a visual inspection was done to remove those galaxies in which GELATO associated a broad component due to the noise near the [OIII]$\lambda5007$\AA\ line. Since the final model is the one with the lowest AIC, a more complex model might be selected unnecessarily. To address this, we checked in how many stacked spectra the [OIII] lines present a better fit with a double Gaussian, rather than a single Gaussian when comparing the $\chi^2$ value of both fits and performing a second visual inspection. If the double Gaussian fit has $\chi^2\leq 20$ and it improves the $\chi^2$ of the single fitting  by at least 20$\%$, the galaxy is included in our sample of outflow candidates (e.g. \citealt{hao2005active}; \citealt{reines2013dwarf}). At this point, the number of AGN dwarf galaxies that fulfil our criteria to show outflows signatures is 64.\\ 

To further restrict our sample of  dwarf galaxies with possible AGN outflows, we performed a similar analysis as the one explained above, but this time for the individual AGN and composite spaxels of the 64 candidates selected. We took advantage of the MaNGA Data Reduction Pipeline (DRP) \footnote{\url{https://www.sdss4.org/dr17/manga/manga-data/data-access/}} of DR17 data-products, which provide  flux-calibrated, sky-subtracted, co-added data cubes from each of the individual exposures for a given galaxy. The wavelength, flux, and ivar of each spaxel are provided by the LOGCUBE data cube. Only spaxels with a S/N $\geq 10$ were selected in order to avoid unreliable spectrum fittings (Appendix \ref{app:A} offers a discussion of the S/N threshold selection can be found). After running GELATO for each of the individual spaxels of the 64 dwarf galaxies with AGN outflow candidates, we selected only those galaxies that present more than ten consecutive AGN spaxels with a broad component in the [OIII] line ($n_\mathrm{out} > 10$) in at least one region of the host galaxy. By applying this cut, a final sample of 13 dwarf galaxies with AGN outflow candidates was obtained, of which seven were classified as AGN galaxies and six as SF-AGN, according to \cite{mezcua2024manga}. Three of the AGN dwarf galaxies (8442-1901, 8982-3703 and 9889-1902) show additional evidence to support the presence of AGN based on the detection of radio emission and one (10223-3702) based on its mid-infrared colours, according to the criteria of \cite{jarrett2011spitzer} or \cite{stern2012mid}. The analysis of the individual spaxels is also performed for the SF spaxels of these 13 dwarf galaxies in order to discuss the origin of the outflow (see Sect. \ref{subsection:origin}).

\subsection{Star formation rate measurement}
\label{subsection:SED}
We fit the spectral energy distributions (SED) using Code Investigating GALaxy EMission (CIGALE; \citealt{boquien2019cigale}) to estimate the SFR of the 13 MaNGA dwarf galaxies with AGN outflow candidates; this is because the one obtained by taking the H$\alpha$ emission line from the MaNGA data-analysis pipeline (DAP; \citealt{westfall2019data}; \citealt{belfiore2019data}) does not account for the AGN contribution (see Sect. \ref{subsection:SNe_outflows}). We considered the single stellar population models
 from \cite{bruzual2003stellar}, assuming a Chabrier initial mass function as per \cite{chabrier2003galactic}. We used the standard nebular emission model from \cite{inoue2011rest} and the dust attenuation model using the \cite{calzetti2000dust} starburst attenuation curve. The reprocessed dust emission was modelled using the \cite{draine2013andromeda} dust models. Finally, we used the AGN emission model from \cite{fritz2006revisiting}. When available, we used the following photometry: SDSS magnitudes in the optical/infrared bands (u,g,r,i,z; \citealt{fukugita1996sloan}; \citealt{doi2010photometric}) adapted to a composite model following \cite{maraston2013stellar}. Near-infrared photometry in the J (1.25$\mu m $), H (1.65$\mu m $), and K (2.17$\mu m$) bands from The Two Micron All Sky Survey (2MASS; \citealt{cutri2003vizier, cutri2012vizier}; \citealt{skrutskie2006two}). Infrared photometry from the Wide-field Infrared Survey Explorer (WISE; \citealt{wright2010wide}; \citealt{cutri2021vizier}) passbands W1, W2, W3, and W4 with effective wavelengths of 3.4, 4.6, 12.1, and 22.5 $\mu m$. Finally, the far- and near-ultraviolet measurements are also included from the Galaxy Evolution Explorer (GALEX;\citealt{bianchi2011galex,bianchi2017revised}; \citealt{osborne2023improved}).

\subsection{Sample of massive galaxies in MaNGA}
To compare our results with a sample of massive galaxies, we used the MaNGA sample of massive galaxies ($M_\ast > 10^{10} M_\odot$) with AGN outflows of \cite{wylezalek2020ionized}. To ensure that the analysis is carried out in the same way for both dwarf and massive galaxies, we performed the same analysis as for the dwarf galaxies described in Sect. \ref{section:dwarfsample} for the 154 massive galaxies with AGN from \cite{wylezalek2020ionized}. We found that  46 out of these 154 massive galaxies present an extended broad component region (ten consecutive AGN spaxels with a broad component in the [OIII] line). \cite{wylezalek2020ionized}, considering only those AGN galaxies with W$_{80}\geq 500$ km s$^{-1}$ (where W$_{80}$ is the velocity width containing 80$\%$ of the flux of the emission line), reported 29 AGN outflows in their sample of massive galaxies. This cut in the W$_{80}$ velocity is the main reason behind the difference between the number of outflows we found in their sample and those reported in \cite{wylezalek2020ionized}. From the 46 massive galaxies with AGN outflow signatures we report here, 17 have W$_{80}\geq 500$ km s$^{-1}$, with  11 out of these 17 shown to be consistent with the 29 reported by \cite{wylezalek2020ionized}. For the remaining 18 galaxies with W$_{80}\geq 500$ km s$^{-1}$ reported by \cite{wylezalek2020ionized}, we found (according to our measurements; see Sect. \ref{section:out_vel}) that 6 of them have W$_{80}\leq 500$ km s$^{-1}$, while the rest do not meet our criteria regarding consecutive AGN spaxels with a broad component in the [OIII] emission line. \\



\section{Results and discussion}
\label{section:results}

In this work, we report the discovery of 13 MaNGA dwarf galaxies with AGN outflow candidates identified using the MaNGA survey. Furthermore, we investigate whether the gas kinematics differs from that of the stellar kinematics. Although gas and stars can rotate perpendicularly (\citealt{sarzi2006sauron}), the misalignment may suggest past interactions (\citealt{casanueva2022origin}; \citealt{raimundo2023increase}; \citealt{zinchenko2023gas}; \citealt{winiarska2025}) or the presence of potential outflows (\citealt{ristea2022sami}). To this end, the  position angle (PA) of the [OIII] total gas velocity and of the stellar velocity can be derived using the PaFit package in Python (\citealt{krajnovic2006kinemetry}; see Table \ref{tab:PA}). For nine out of the 13 dwarf galaxies with outflows signatures, we found that the $|\mathrm{PA}_{\mathrm{gas}}-\mathrm{PA}_{\mathrm{stars}}|\gtrsim20\degree$. This supports the assumption of a deviation of the gas from the stellar kinematics (see Figs. \ref{fig:velmaps} and \ref{fig:velmapsext}).  The remaining four galaxies, 7992-6102, 8442-1901, 8657-6104, and 9889-1902 (which is in the process of undergoing a merger), have a PA difference of $\leq 12\degree$, which does not confirm the outflowing nature of the gas. \\

In the following subsections, we present the outflow kinematic and energetic properties of the 13 MaNGA dwarf galaxies. We start with the calculation of the velocity of the outflows, followed by calculations of the radius, the mass and the energy,  momentum, and mass rates. 

\subsection{Outflow velocity and radius}
\label{section:out_vel}
The outflow velocity is defined in \cite{manzano2019agn} as

\begin{equation}
\label{eqn:vout}
\mathrm{v}_\mathrm{out}=-\mathrm{v}_0+\frac{W_{80}}{2}
,\end{equation}

\noindent where $\mathrm{v}_0$ is the velocity offset between the narrow and the broad component of the [OIII]$\lambda$5007\AA\ line. In the case of a single Gaussian fitting, W$_{80}$ could be defined as W$_{80}=$ 1.09 $\times$ FWHM. However, for non-Gaussian emission line profiles, the non-parametric velocity width measurements are more sensitive to the weak broad bases \citep{liu2013observations}. As our fitting is composed by the sum of two Gaussians, W$_{80}$ is calculated using the following expression:

\begin{table}
\centering
    \caption{Position angles.}
    \begin{tabular}{cccc}
    \hline
    MaNGA & PA$_{\mathrm{gas}}$ & PA$_{\mathrm{stars}}$ & $|\mathrm{PA}_{\mathrm{gas}}-\mathrm{PA}_{\mathrm{stars}}|$\\
    plateifu & ($\degree$) & ($\degree$) & ($\degree$)\\
    (1) & (2) & (3) & (4)\\ \hline \hline
        10223-3702 & 112  & 143 & 31\\ 
        10226-1901 & 118 & 56 & 62\\
        11022-12702 & 124  & 149 & 25 \\
        11754-3701 & 44  & 25 & 19  \\         
        11826-12702 & 119 & 142 & 25\\ 
        12487-3704 & 149 & 62 & 87 \\ 
        7992-6102 & 142  & 136 & 6\\ 
        8252-1902 & 112  & 62 & 50\\  
        8442-1901 & 86  & 74 & 12\\  
        8655-6101 & 93 &  68 & 25 \\ 
        8657-6104 & 12 & 12 & 0\\ 
        8982-3703 & 56 & 6 & 50\\ 
        9889-1902 & 118 & 118 & 0 \\ \hline

    \end{tabular}
    \tablefoot{(1) MaNGA plateifu; (2) position angle of the [OIII] gas; (3) position angle of the stellar component; (4) difference between the gas and stellar position angle. }
\label{tab:PA}
\end{table}

\begin{equation}
W_{80}\equiv \mathrm{v}_{90}-\mathrm{v}_{10}
,\end{equation}

\noindent where $\mathrm{v}_{90}$ and $\mathrm{v}_{10}$ are the velocities at the 10th and 90th percentiles of the total flux. In order to calculate these velocities, we need to calculate the cumulative flux as a function of the velocity:

\begin{figure*}
\centering
\includegraphics[width=\textwidth]{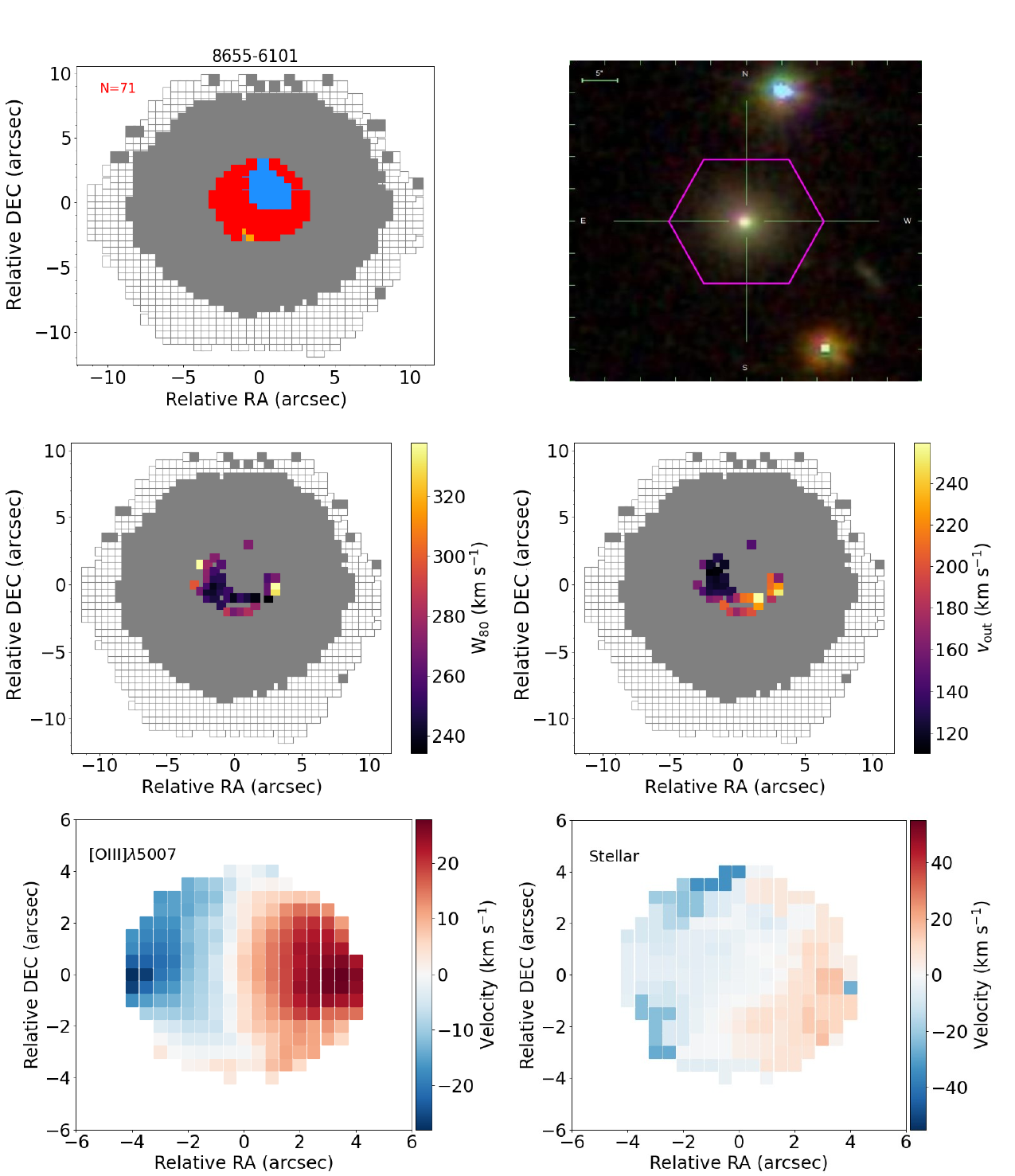} 
\caption{MaNGA analysis for one of the dwarf galaxies with AGN outflow candidates (8655-6101). Top left: Spatial distribution of the BPT-classified spaxels (red: AGN; green: composite; blue: SF; orange: LINER). The `N' shows the number of AGN and composite spaxels used in the analysis. Top right: SDSS composite image. The pink hexagon shows the IFU coverage. Middle: Spatial W$_{80}$ (left) and $\mathrm{v}_\mathrm{out}$ (right) outflow velocity distribution, where we can see the extension of the outflow. Empty squares mark the IFU coverage and grey squares those spaxels with a continuum S/N $> 1$. Bottom: MaNGA [OIII]$\lambda$5007\AA\ velocity map (left) and stellar velocity (right).}
\label{fig:velmaps}
\end{figure*}

\begin{figure*}
\centering
\includegraphics[width=\textwidth]{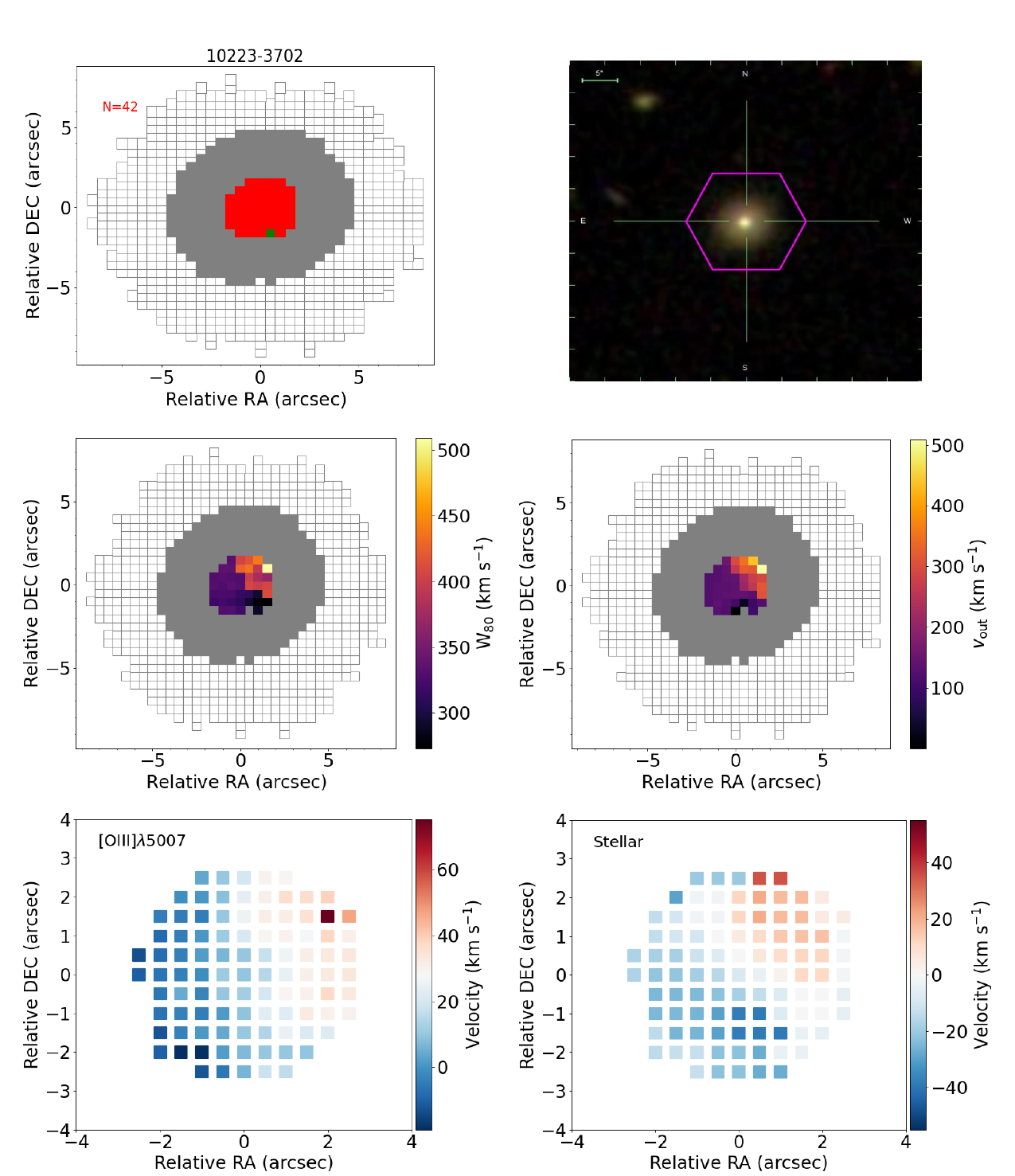} 
\caption{Same as Fig. \ref{fig:velmaps} but for the dwarf galaxy 10223-3702.}\label{fig:velmapsext}
\end{figure*}
\begin{equation}
\label{eqn:integral}
\phi(\mathrm{v})\equiv \int_{-\infty}^\mathrm{v} F_\mathrm{v}(\mathrm{v}')d\mathrm{v}'
.\end{equation}

The total line flux is given by $\phi(\infty)$. $F_\mathrm{v}$ is the spectral flux density. In the wavelength space, the spectral flux density can be parametrised as a Gaussian distribution:

\begin{equation}
F_\lambda=\frac{F}{\sigma_\lambda\sqrt{2\pi}}\mathrm{exp}\left(-\frac{1}{2}\left(\frac{\lambda-\lambda_\mathrm{r}(1+z)}{\sigma_\lambda}\right)^2\right)
,\end{equation} 

\noindent taking into account the expansion of the Universe with the factor ($1+z$). Then, $F$ is the integrated line flux, $\lambda$ is the wavelength, $\lambda_\mathrm{r}$ is the wavelength of the emission line in the rest-frame, and $\sigma_\lambda$ is the dispersion in units of wavelength. As the dispersion of the line is dominated by the kinematics of the gas, we convert $\sigma_\lambda$ into velocity units by using the Doppler effect:

\begin{equation}
\sigma=c\frac{\sigma_\lambda}{\lambda_\mathrm{r}(1+z)}
.\end{equation} 



\noindent We notice that the integrand in Eq. \ref{eqn:integral} is in the velocity space. By definition, $F_\mathrm{v} d\mathrm{v}=F_\lambda d\lambda$, meaning that $\int F_\lambda d\lambda =\int F_\mathrm{v} d\mathrm{v}$, as long as the bounds of integration are properly adjusted. Since $1+z=1+\mathrm{v}/c=\lambda/\lambda_\mathrm{r}$, it implies $d\lambda/d\mathrm{v}=\lambda_\mathrm{r}/c$. \\

In the case of a double Gaussian fitting, $F_\mathrm{v}$ will be the sum of both Gaussian components: the narrow and the broad. GELATO outputs $\sigma$ and $F$, so we can calculate W$_{80}$ and $\mathrm{v}_\mathrm{out}$. A map illustrating the spatial distribution of the outflow velocity (in terms of W$_{80}$ and $\mathrm{v}_\mathrm{out}$) is shown for two of the 13 dwarf galaxies with AGN outflow candidates (Figs. \ref{fig:velmaps} and \ref{fig:velmapsext}). In Table \ref{tab:summary}, the results are summarised by taking the median of all the AGN and composite spaxels with outflow signatures.\\

There are two different scenarios for calculating the outflow radius: when the full extent of the outflow is not observed (Fig. \ref{fig:velmaps}) and when it is observed (Fig. \ref{fig:velmapsext}). For each of these two cases, a different criteria is adopted:

\begin{itemize}
    \item If it is fully extended: we consider that the diameter is  the largest distance of consecutive AGN and composite spaxels with outflow signatures.
    \item If it is not fully extended: We consider the radius as the distance from the central spaxel to the further AGN or composite spaxel with outflow signature that is in a region with at least ten consecutive AGN or composite spaxels with outflow signatures.
\end{itemize}

As each spaxel is 0.5 arcsec size, the angular size of the outflow can be calculated. Then, through parallax, the projected radius is obtained:

\begin{equation}
R_\mathrm{out}=\mathrm{tan}(\alpha) \ d_\mathrm{l}
,\end{equation}

\noindent where $\alpha$ is half of the angular size in radians and $d_\mathrm{l}$ is the luminosity distance. We use the projected radius as a lower limit of the outflow radius, since the physical size depends on the inclination angle of the outflow with respect to the line of sight. \\

\subsection{Energetics of the outflow}

The ionised gas mass of the outflow can be calculated based on the luminosity of the H$\alpha$ emission line. Considering the sum of all the AGN and composite spaxels with outflow signatures and the equation defined by \cite{osterbrock2006astrophysics}, we have

\begin{equation}
M_\mathrm{out}=\sum_{i=1}^{n_\mathrm{out}} m_\mathrm{out}=\sum_{i=1}^{n_\mathrm{out}} 4.48M_\odot\left(\frac{L_{\mathrm{H}\alpha}}{10^{35}\mathrm{erg} \ s^{-1}}\right)\left(\frac{<n_e>}{1000\mathrm{cm}^{-3}}\right)^{-1}
,\end{equation}

\noindent where $M_\mathrm{out}$ is the total mass of the outflow, $m_\mathrm{out}$ is the outflow mass within individual spaxels, and $L_{H\alpha}$ is the $H\alpha$ luminosity of the individual outflow spaxels. Focusing on those spaxels with outflow, the electron density, $n_e$ is estimated using the [SII]$\lambda$6716\AA\//[SII]$\lambda$6732\AA\ relation \citep{sanders2015mosdef}:\\

\begin{equation}
n_e(R)=\frac{cR-ab}{a-R}
,\end{equation}

\noindent where R is the ratio between the [SII] emission lines flux and a, b and c are dimensionless constants equal to 0.4314, 2.107 and 627.1 respectively. The electron density ranges from 825 cm$^{-3}<n_e<1071$ cm$^{-3}$.\\

As the detected outflows are spatially resolved, the mass (dM/dt), momentum (dP/dt), and kinetic energy (dE/dt) rates can be calculated using the following equations (e.g. \citealt{liu2020integral}; \citealt{bohn2021near}):

\begin{equation}
dM/dt=\sum_{i=1}^{n_\mathrm{out}} dm/dt=\sum_{i=1}^{n_\mathrm{out}} \frac{m_\mathrm{out}\mathrm{v}_0\mathrm{sec}\theta}{R_\mathrm{out}}
,\end{equation}

\begin{equation}
dP/dt=\sum_{i=1}^{n_\mathrm{out}} (\mathrm{v}_0\mathrm{sec}\theta)dm/dt
,\end{equation}

\begin{equation}
dE/dt=\frac{1}{2}\sum_{i=1}^{n_\mathrm{out}} \left[(\mathrm{v}_0 \ \mathrm{sec}\theta)^2+3\sigma_\mathrm{out}^2\right]dm/dt
,\end{equation}

\noindent where $\mathrm{v}_0$ here is the absolute value of the velocity offset between the narrow and broad component (see Eq. \ref{eqn:vout}), $\sigma_\mathrm{out}$ is the velocity dispersion measured from the outflow component within individual spaxels, and $\theta=\mathrm{sin}^{-1}(r_\mathrm{spaxel}/R_\mathrm{out})$, where $r_\mathrm{spaxel}$ is the angular size of the spaxel converted into physical distance. The results of the mass, momentum and kinetic energy rate range from $10^{-4.4}$ < dM/dt (M$_\odot\mathrm{yr}^{-1}$) < 10$^{-1.2}$, from $10^{4.3}$ < c dP/dt ($L_\odot$) < $10^{8.6}$, and from $10^{35.4}$ < dE/dt ($\mathrm{erg} \ \mathrm{s}^{-1}$) < $10^{39.5}$. These results are also listed in Table \ref{tab:summary}.

\begin{figure}
\centering
\includegraphics[width=9cm]{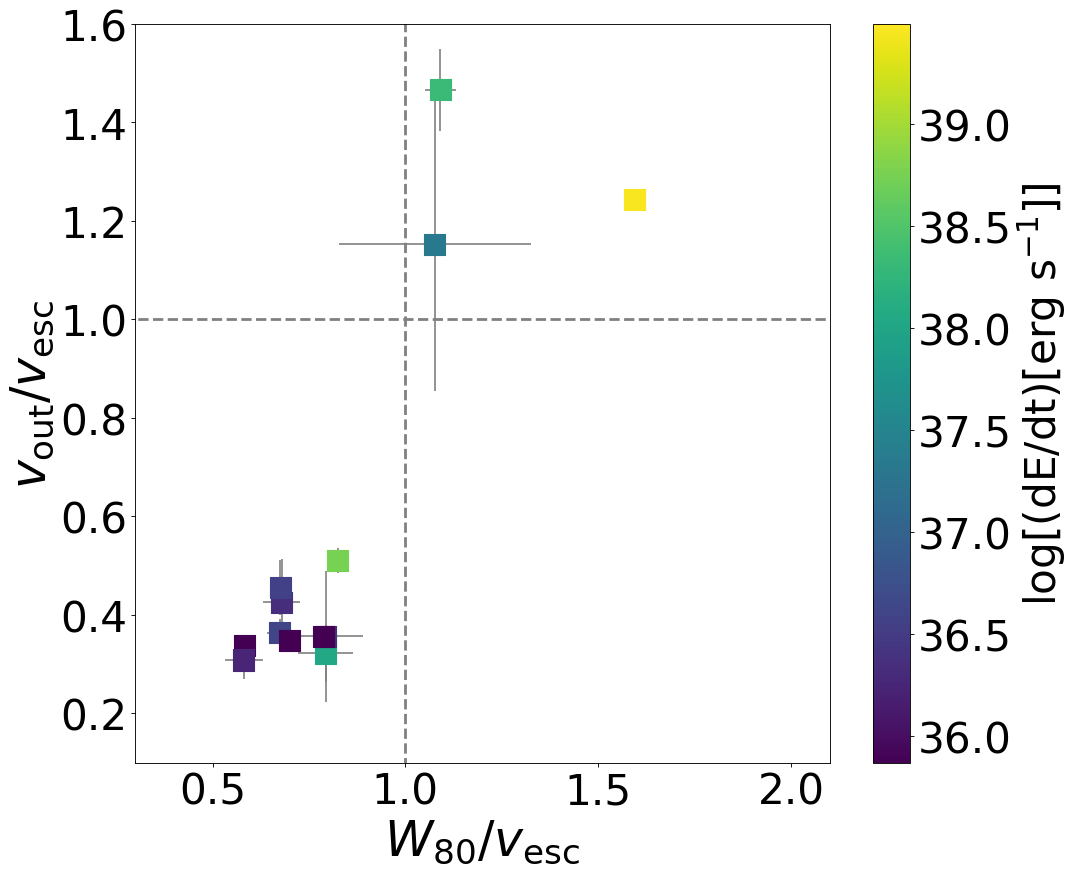} 
\caption{Comparison of the ratios of the two velocity indicators W$_{80}$ and $\mathrm{v}_\mathrm{out}$ with $\mathrm{v}_\mathrm{esc}$. The grey dashed lines represent the 1:1 relation. Three out of the 13 MaNGA dwarf galaxies with AGN outflow candidates would have outflows fast enough to become unbound from the galaxy and from its dark matter halo.}   
\label{fig:vesc}
\end{figure}

\subsection{Outflows escaping their host halos}
\label{Section:v_esc}
To investigate whether the outflows are able to escape the gravitational potential of their host halos, the velocity of the outflows and the escape velocity, $\mathrm{v}_\mathrm{esc}$, are compared. Assuming a Navarro-Frenk-White profile (NFW; \citealt{navarro1997universal}) and making use of the abundance matching (\citealt{moster2010constraints, moster2013galactic}), we can infer the halo mass from the stellar mass of the galaxy:\\

\begin{equation}
    \frac{M_\ast}{M_h}=2\left(\frac{M_\ast}{M_h}\right)_{0}\left[\left(\frac{M_h}{M_1}\right)^{-\beta}+\left(\frac{M_h}{M_1}\right)^{\gamma}\right]^{-1}
.\end{equation}

\noindent This expression has four parameters: the normalisation of the stellar-to-halo mass ratio $(M_\ast/M_h)_0=0.02817$, a characteristic mass of $M_1=10^{11.899}$, and $\beta=0.611$, $\gamma=1.068$, with these latter two slopes serving as an indication of the behaviour of the mass ratio at the low-mass and high-mass ends, respectively. Then, a gravitational potential is considered at the centre $\phi(r=0)$ from a spherical NFW profile (\citealt{lokas2001properties}): 

\begin{equation}
    \phi(r=0)=-\frac{4}{3}cg(c)\pi Gr_v^2 v\rho_c^0
,\end{equation}

\noindent where $c=10$ is the concentration parameter, $g(c)=1/(\mathrm{ln}(1+c)-c/(1+c))$ is a function of the concentration parameter, $v=200$ is the virial overdensity, $\rho_c^0= 277.5 M_\odot /\mathrm{kpc}^3$ is the present critical density, and $r_v=(3M_h/4\pi v \rho_c^0)^{1/3}$ is the virial radius. Finally, using the expression defined in \cite{manzano2019agn} we obtain the escape velocity:

\begin{equation}
    \mathrm{v}_\mathrm{esc}^2=2|\phi(r=0)|
.\end{equation}

In Fig. \ref{fig:vesc} the velocity of the outflow is compared with $\mathrm{v}_\mathrm{esc}$. While W$_{80}$ is the most common value when defining the outflow velocity, $\mathrm{v}_\mathrm{out}$ has recently been also adopted as outflow velocity estimator (e.g. \citealt{manzano2019agn}; \citealt{bohn2021near};   \citealt{zheng2023escaping}). According to both W$_{80}$ and $\mathrm{v}_\mathrm{out}$, 8252-1902, 8442-1901 and 8982-3703 would have outflows consistent with the high enough velocity needed to escape the galaxy and the dark matter halo. Also, in Fig. \ref{fig:vesc}, we show a colourbar with the kinetic energy rate of the galaxies. The three dwarf galaxies whose outflows are able to escape are among the five with the highest kinetic energy rate. For these three galaxies, the outflowing gas could contribute to AGN feedback by removing gas from the central region or could cause feedback to the galactic halo by heating the intergalactic medium (\citealt{smethurst2021kiloparsec}).\\

\subsection{Driver behind  these outflows}
\label{subsection:origin}
Although we have primarily based our analysis on the selection of AGN spaxels to detect AGN outflows, it was necessary to engage in a deeper study in order to understand the origin of the outflows detected.\\

\subsubsection{Considering stellar winds from massive stars or SNe as a potential driver }
\label{subsection:SNe_outflows}
Massive stars are able to inject energy and momentum to the interstellar medium  through outflows coming from stellar winds and SNe explosions. These stellar processes are commonly believed to be the main source of feedback in dwarf galaxies (e.g. \citealt{veilleux2005galactic}; \citealt{martin2018exploring}). In this section, we investigate whether the 13 MaNGA dwarf galaxies with outflows could be driven by stellar processes by computing the expected energy driven by stellar wind outflows in massive stars or by SNe outflows.\\

 For stellar winds in massive stars, the kinetic energy rate can be derived as $dE/dt=\frac{1}{2}dM/dt \  \mathrm{v}_{\mathrm{out}}^2$ (\citealt{rosen2022massive}), where $dM/dt$ is the mass rate, which typically ranges from $10^{-5}$ to $10^{-4}$ $M_{\odot}/\mathrm{yr}$ (\citealt{bally2008outflows}). The outflow velocity in massive stars rarely surpasses 100 km s$^{-1}$ (e.g. \citealt{churchwell1997origin}; \citealt{commerccon2022discs}). Consequently, the kinetic energy rate is expected to be in the range of $10^{34-35}$ erg s$^{-1}$, which is lower than our results. Thus, we can safely rule out massive stars from the plausible driving mechanisms.\\

For SNe, the kinetic energy rate can be calculated as $\sim 7\times 10^{41}(\mathrm{SFR}/M_\odot \ \mathrm{yr}^{-1})$ (\citealt{veilleux2005galactic}). We calculated the SFR by taking the total H$\alpha$ and H$\beta$ emission lines from the MaNGA DAP and using the equation $\mathrm{log(SFR)=log}(L_{H_\alpha})- 41.27$ (\citealt{kennicutt2012star}), after having previously corrected $L_{\mathrm{H}\alpha}$ from extinction using the \cite{calzetti2000dust} extinction law. A range of SNe kinetic energy rate that goes from dE/dt$\sim 10^{40}$ erg s$^{-1}$ to $\sim 10^{42}$ erg s$^{-1}$ is obtained. The expected SNe kinetic energy rate is one to six orders of magnitude higher than the kinetic energy rate of the 13 MaNGA dwarf galaxies with outflows. Consequently, in terms of energy, the 13 outflows detected could be driven by SNe as the kinetic energy rate of the SNe is able to explain the presence of outflows up to $10^{40-42}$ erg s$^{-1}$. However, the SFR might be overestimated in those cases where the AGN dominates (see Fig. \ref{fig:velmapsext}). This comes from the fact that the H$\alpha$ emission line comes mostly from the AGN. To overcome this problem, we derived the SFR from the SED fitting (see Sect. \ref{subsection:SED}). The SFR obtained from the SED fitting is similar or even higher with those obtained from the H$\alpha$ emission line, which is in agreement with the results obtained from \cite{siudek2024value}. Alternatively, we also used a SFR indicator based on the [OII]$\lambda3726$ emission line (SFR(OII); e.g. \citealt{kennicutt1998star}; \citealt{hopkins2003star}; \citealt{kewley2004ii}). We corrected the [OII]$\lambda3726$ emission line for extinction following \cite{calzetti2000dust} extinction law and correct for AGN contribution by subtracting ten per cent of the total [OIII]$\lambda5007$ luminosity (\citealt{vietri2022type}). The SFR(OII) obtained is typically one order of magnitude lower than that obtained from the H$\alpha$ emission line and from the SED fitting. However, the SNe kinetic energy rate computed from the SFR(OII) still remains several orders of magnitude higher than the kinetic energy rate of the 13 MaNGA dwarf galaxies with outflows. \\

\begin{figure}
\centering
\includegraphics[width=9cm]{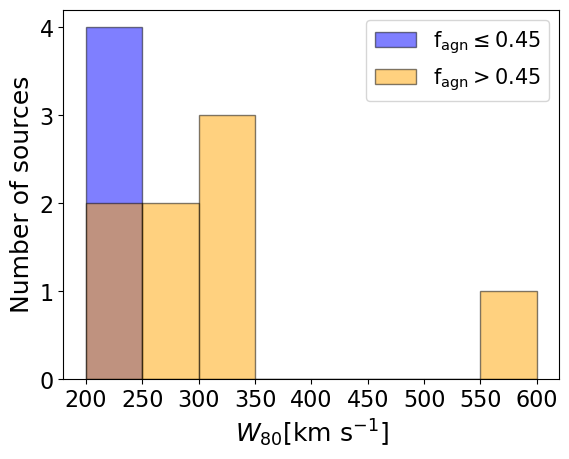} 
\caption{Distribution of the W$_{80}$ velocity for the dwarf galaxies with AGN outflow candidates with AGN fraction $\leq 0.45$ (blue), and with AGN fraction $>0.45$ (orange).}   
\label{fig:hist}
\end{figure}
\subsubsection{AGN fraction and W$_{80}$ velocity}

In this section, we look at the AGN fraction, f$_{\mathrm{AGN}}=n_{\mathrm{agn}}/(n_{\mathrm{agn}}+n_{\mathrm{SF}})$, with $n_{\mathrm{agn}}$ as the number of spaxels classified as AGN and $n_{\mathrm{SF}}$ the number of spaxels classified as SF. A Mann-Whitney U test was performed to check whether the W$_{80}$ velocity of the 12 dwarf galaxies with AGN outflow candidates (8442-1901 is excluded as all the spaxels with S/N $\geq 10$ are composite\footnote{8442-1901 is however classfied as an AGN by \cite{mezcua2024manga} as they consider spaxels with a
S/N $\geq$ 3.}, see Sect. \ref{section:dwarfsample}) comes from different distributions in cases where the AGN fraction is taken into account. We find that when the AGN fraction of > 0.45, while the W$_{80}$ velocities are drawn from a different distribution at a $\sim99$\% confidence level, with a p-value of $\sim$0.004 (see Fig. \ref{fig:hist}). These velocities are higher (from 231 km s$^{-1}$ to 566 km s$^{-1}$) than those corresponding to the galaxies with an AGN fraction < 0.45 (from 205 km s$^{-1}$ to 227 km s$^{-1}$). Eight out of the 12 MaNGA dwarf galaxies have an AGN fraction > 0.45, six of them are classified as AGN, and two as SF-AGN (see Table \ref{tab:summary}). Although the galaxy 8442-1901 was not included in this analysis, its W$_{80}$ velocity (396 km s$^{-1}$) is consistent with those from the eight galaxies with AGN fraction > 0.45.  Therefore, since AGN outflows are expected to be faster than those driven by stellar processes (\citealt{liu2020integral}; \citealt{wylezalek2020ionized}), the AGN emerges as the main candidate to act as the driving mechanism of the outflows observed in these nine galaxies.\\



\subsubsection{Study of the SF spaxels}

We also studied the detection of outflows in the SF spaxels of our sample of 13 dwarf galaxies following the same procedure for the individual spaxels as that described in Sect \ref{section:dwarfsample}. We excluded the galaxy 8442-1901 as it presents only composite spaxels. For the six dwarf galaxies classified as SF-AGN (see Table \ref{tab:summary}), the W$_{80}$ is consistent within the errors when comparing the SF spaxels against the AGN spaxels. On the contrary, For the six dwarf galaxies classified as AGN, four of them (10223-3702, 12487-3704, 8252-1902 and 8982-3703) have an AGN fraction f$_{\mathrm{AGN}}\sim1$ so, although we were not able to carry out this analysis, the absence of spaxels ionised by SF suggests that for these three galaxies the AGN must drive the outflow. The two remaining galaxies (8655-6102 and 9889-1902) present lower W$_{80}$ velocities in the SF spaxels than in the AGN spaxels, even when accounting for their errors. This indicates that for these two galaxies the AGN could be the main candidate for driving the outflows. \\

Regarding the kinetic energy rate we find consistent values, within the errors, for those obtained from SF spaxels compared to those obtained from AGN spaxels. However, for the galaxy 9889-1902, the kinetic energy rate calculated from the AGN spaxels is one order of magnitude higher. This makes 9889-1902 a strong candidate for hosting AGN outflows, as they are expected to be more energetic than SF outflows (\citealt{aravindan2023comparison}).

\subsubsection{AGN bolometric luminosity and kinetic energy rate}
\label{section:drivelum}

The AGN bolometric luminosity (L$_{\mathrm{bol}}$) is given by L$_{\mathrm{bol}}=1000\times\mathrm{L_{[OIII]}}$ (\citealt{moran2014black}), where $\mathrm{L_{[OIII]}}$ is the sum of the [OIII]$\lambda$5007 luminosities of those spaxels classified as AGN or composite in the BPT diagrams (\citealt{mezcua2024manga}). The AGN bolometric luminosity L$_{\mathrm{bol}}$ for the 13 MaNGA dwarf galaxies with AGN outflow candidates (see Table \ref{tab:kineticrate}), is four to six orders of magnitude higher than the kinetic energy rate of the outflow, implying that the AGNs in these 13 galaxies are capable of driving these outflows (\citealt{liu2020integral}). \\

In summary, for the seven MaNGA dwarf galaxies classified as AGN, six are more likely to have outflows driven by the AGN, as indicated by their AGN fraction being higher than that of the six dwarf galaxies classified as SF-AGNs (with their L$_{\mathrm{bol}}$ being greater than the kinetic energy rate and a higher W$_{80}$ velocity in the AGN spaxels than in the SF spaxels). For the remaining dwarf galaxy classified as an AGN (8442-1901), the outflow is also likely to be driven by the AGN based on its high W$_{80}$ velocity and its bolometric luminosity. In the case of the six MaNGA dwarf galaxies classified as SF-AGNs, two (10226-1901 and 11754-3701) have an AGN fraction and W$_{80}$ value  consistent with their having been drawn from the same distribution as the five AGN galaxies. For the remaining four SF-AGNs, it is difficult to distinguish the driving mechanism of the outflow, therefore both SNe and AGNs are possible candidates for driving these outflows. \\

At this point, we divided the sample of 13 MaNGA dwarf galaxies with AGN outflow candidates in two sub-samples: nine MaNGA dwarf galaxies consistent with AGN outflows (the seven AGNs and two SF-AGNs: 10226 and 11754-3701) and four MaNGA dwarf galaxies with uncertain outflows (the remaining four SF-AGNs). We emphasise that for all the 13 dwarf galaxies we cannot rule out the possibility of SNe as the driving mechanism in this scenario, based on the fact that the SNe kinetic energy rate exceeds the outflow kinetic energy rate in all cases.

\subsection{Mass-loading factor}
\label{section:massloading}
The mass-loading factor is a parameter used to compare the amount of gas that is ejected by the outflow to that consumed by SF (\citealt{harrison2024observational}). It is defined as the ratio of the outflow mass rate compared with the SFR, $\eta=(dM/dt)/\mathrm{SFR}$. In the MaNGA sample of nine dwarf galaxies with AGN outflows, we find that the mass-loading factor for all of them is $\eta\ll1$ for all three SFR indicators: the one obtained from the H$\alpha$ emission line, the one derived from SED fitting, and the SFR(OII). This implies that the amount of gas used for SF processes is greater than the one ejected by the outflow, therefore these outflows are unlikely to quench the SF in the host galaxy (\citealt{carniani2024jades}). However, the mass-loading factor can be misleading due to the fact that the spatial scale when calculating the energetic properties of the outflows and the scale needed to obtain the integrated SFR are different; namely, scales of a few kiloparsecs  versus galaxy scales (\citealt{harrison2024observational}), except for the galaxies 11754-3701 and 11826-12702,  where $R_\mathrm{out}>5$ kpc.

\subsection{Comparison with other works on dwarf galaxies}
\label{section:w80}

\begin{table*}
    \caption{Properties of the outflows for the MaNGA sample of 13 dwarf galaxies. }

    \begin{adjustbox}{width=1\textwidth}
    \begin{tabular}{ccccccccccccccc}
    \hline
    MaNGA & RA & DEC & z & log($M_\ast$)& SFR & f$_{\mathrm{AGN}}$ & W$_{80}$ & $\mathrm{v}_\mathrm{out}$ & $R_\mathrm{out}$ & log($M_\mathrm{out}$) & log(dM/dt) & log(dE/dt)& log(c dP/dt)& Type\\
    plateifu & (J2000) & (J2000) &  & (M$_{\odot}$)& (M$_{\odot}$ yr$^{-1}$) & & (km s$^{-1}$) & (km s$^{-1}$) & (kpc) & (M$_{\odot}$) & (M$_{\odot}$ yr$^{-1}$) &(erg s$^{-1}$) &($L_\odot$)&  \\
    (1) & (2) & (3) & (4) & (5) & (6) & (7)& (8)&(9)&(10)&(11)&(12)&(13)& (14) & (15) \\ \hline \hline
        10223-3702 & 33.524643 & -0.276961 & 0.0373 & 9.8 & 0.18 & 1.00& 330 $\pm$ 30 & 134 $\pm$ 24 & 1.3 $\pm$ 0.4 & 5.00 $\pm$ 0.02 & -2.38 $\pm$ 0.04 & 38.0 $\pm$ 0.6 & 7.30 $\pm$ 0.08 & AGN$^{\dagger}$ \\
        10226-1901* & 38.208904 & -0.353272 & 0.0204 & 9.0 & 0.15 & 0.63 &238 $\pm$ 29 & 107 $\pm$ 40 & 1.5 $\pm$ 0.3  & 4.09 $\pm$ 0.01 & -3.7 $\pm$ 0.1 & 36.4  $\pm$ 0.7 & 5.9 $\pm$ 0.1 & SF-AGN \\ 
        11022-12702 & 219.914586 & 5.101727 & 0.0228 & 8.9 & 0.17 & 0.44 & 205 $\pm$ 2 & 102 $\pm$ 5 & 0.8 $\pm$ 0.7 & 4.04 $\pm$ 0.02 & -4.3 $\pm$ 0.4 & 35.5 $\pm$ 0.7 & 4.3 $\pm$ 0.3 & SF-AGN \\
        11754-3701* & 129.710888 & 1.405020 & 0.0290 & 9.3 &  1.16 & 0.51 &231 $\pm$ 6 & 156 $\pm$ 19 & 5.2 $\pm$ 0.5 & 4.50 $\pm$ 0.01 & -3.54 $\pm$ 0.04 & 36.6 $\pm$ 0.5 & 6.1 $\pm$ 0.1 & SF-AGN \\
        11826-12702 & 188.522918 & 36.841747 & 0.0402 & 9.7 & 0.74 & 0.34 &227 $\pm$ 6 & 132 $\pm$ 8 & 4.8 $\pm$ 0.5  & 4.16 $\pm$ 0.01 & -4.24  $\pm$ 0.04 & 35.65 $\pm$ 0.06 & 4.78 $\pm$ 0.04 & SF-AGN \\ 
        12487-3704 & 139.326370 & 29.727896 & 0.0341 & 9.6 & 0.16 & 1.00 &261 $\pm$ 13 & 141 $\pm$ 11  & 1.9 $\pm$ 0.4  & 4.65 $\pm$ 0.02 & -3.37 $\pm$ 0.05 & 36.61 $\pm$ 0.07 & 5.76 $\pm$ 0.06 & AGN \\ 
        7992-6102* & 253.889080 & 63.242118 & 0.0225 & 9.6 & 1.63 & 0.30 & 226 $\pm$ 19 & 120 $\pm$ 15 & 2.1 $\pm$ 0.3 & 4.19  $\pm$ 0.01 & -3.8 $\pm$ 0.1 & 36.3 $\pm$ 1.4 & 5.77 $\pm$ 0.09 & SF-AGN \\ 
        8252-1902 & 146.091838 & 47.459851 & 0.0259 & 9.4 & 0.06 & 0.98 & 325 $\pm$ 75 & 348 $\pm$ 90 & 0.9 $\pm$ 0.5 & 3.54  $\pm$ 0.02 & -3.2 $\pm$ 0.1 & 37.34 $\pm$ 0.06 & 6.84 $\pm$ 0.05 & AGN \\ 
        8442-1901 & 199.675905 & 32.918662 & 0.0358 & 9.7 & 1.92 & - & 396 $\pm$ 14 & 532 $\pm$ 30 & 1.3 $\pm$ 0.3 & 4.19  $\pm$ 0.01 & -2.47 $\pm$ 0.01 & 38.32 $\pm$ 0.01 & 7.74 $\pm$ 0.01 & AGN$^{\dagger\dagger}$\\ 
        8655-6101* & 358.094460 & -0.628760 & 0.0226 & 9.6 & 0.09 & 0.62 &257 $\pm$ 18 & 161 $\pm$ 33 & 1.5 $\pm$ 0.2  & 3.91 $\pm$ 0.02 & -3.71 $\pm$ 0.08 & 36.4 $\pm$ 0.3 & 5.8 $\pm$ 0.1 & AGN \\ 
        8657-6104* & 10.417361 & 0.208729 & 0.0175 & 8.7 & 0.12 & 0.24 &213 $\pm$ 4  & 96 $\pm $ 6 & 0.4 $\pm$ 0.2 & 3.19  $\pm$ 0.04 & -4.43  $\pm$ 0.08 & 35.4 $\pm$ 0.1 & 4.37 $\pm$ 0.08 & SF-AGN \\ 
        8982-3703 & 203.190090 & 26.580376 & 0.0470 & 9.4 & 0.92 & 1.00 & 566 $\pm$ 5  & 441 $\pm$ 5  & 2.2$\pm$ 0.5  & 5.85 $\pm$ 0.01  & -1.28 $\pm$ 0.01  & 39.46 $\pm$ 0.01  & 8.61 $\pm$ 0.01 & AGN$^{\dagger\dagger}$ \\ 
        9889-1902 & 234.858582 & 24.943586 & 0.0228 & 9.7 & 1.11 & 0.78 & 333 $\pm$ 7 & 206 $\pm$ 10 & 2.5 $\pm$ 0.5 & 5.55 $\pm$ 0.01  & -1.77 $\pm$ 0.01  & 38.75 $\pm$ 0.01  & 8.14 $\pm$ 0.01 & AGN$^{\dagger\dagger}$  \\ \hline 
    \end{tabular}
    \end{adjustbox}
    \tablefoot{ (1) MaNGA plateifu; (2,3) RA, DEC coordinates of the optical center of the galaxy or IFU center; (4) galaxy redshift; (5) galaxy stellar mass; (6) star formation rate derived from H$\alpha$; (7) AGN fraction (8) median value of the W$_{80}$ velocity; (9) median value of the velocity of the outflow; (10) radius of the outflow; (11) ionised gas mass of the outflow; (12) ionised gas mass outflow rate; (13) ionised gas kinetic energy outflow rate; (14) ionised gas momentum outflow rate; and (15) MaNGA BPT classification. * Galaxies without fully extended outflows. $^\dagger$ MIR AGN, $^{\dagger\dagger}$ radio AGN (see \citealt{mezcua2024manga}).}
\label{tab:summary}
\end{table*}

In this section, we compare the outflow velocity (as $\mathrm{v}_\mathrm{out}$ and W$_{80}$) and luminosity obtained for the sample of 13 MaNGA dwarf galaxies with AGN outflow candidates with other studies. For the nine MaNGA dwarf galaxies with AGN outflows the W$_{80}$ velocity ranges from 231 to 566 km s$^{-1}$ and $\mathrm{v}_\mathrm{out}$ from 107 to 532 km s$^{-1}$. In the case of 9889-1902 and 7992-6102, their values for W$_{80}$ (333 km s$^{-1}$ and 226 km s$^{-1}$, respectively) and $\mathrm{v}_\mathrm{out}$ (206 km s$^{-1}$ and 120 km s$^{-1}$, respectively) could be enhanced due to the merger process in which they are involved. For the four MaNGA dwarf galaxies with uncertain outflows, the W$_{80}$ values range from 205 to 226 km s$^{-1}$ and $\mathrm{v}_\mathrm{out}$ from 96 to 132 km s$^{-1}$. These results were compared with those of other dwarf galaxies with AGN outflow candidates. By using IFU spectroscopy, \cite{liu2020integral} studied 8 out of the 29  dwarf galaxies previously studied in \cite{manzano2019agn} using longslit spectroscopy. The authors detected AGN outflows with W$_{80}$ ranging 140-980 km s$^{-1}$, considering a [OIII]$\lambda5007$\AA\ line profile with up to three Gaussians components (i.e. C1, C2, and C3), where C2 and C3 show strong evidence of outflows. \cite{bohn2021near} also studied the same sample as \cite{manzano2019agn} but focusing on the [Si VI] 1.9630$\mu$m line instead of the [OIII]$\lambda5007$\AA\ and obtaining an outflow velocity range of 155 $\leq \mathrm{v}_\mathrm{out} \ (\mathrm{km} \ \mathrm{s}^{-1})\leq$ 770 and 108 $\leq$ W$_{80} \ (\mathrm{km} \ \mathrm{s}^{-1})\leq$ 1350. \cite{zheng2023escaping} found AGN outflows with a velocity of $\mathrm{v}_\mathrm{out}=471$ km s$^{-1}$ for SDSS J0228-0901. The velocities of the 13 MaNGA dwarf galaxies with AGN outflow candidates are therefore consistent with those previously studied. \\

\begin{figure}
\centering
\includegraphics[width=9cm]{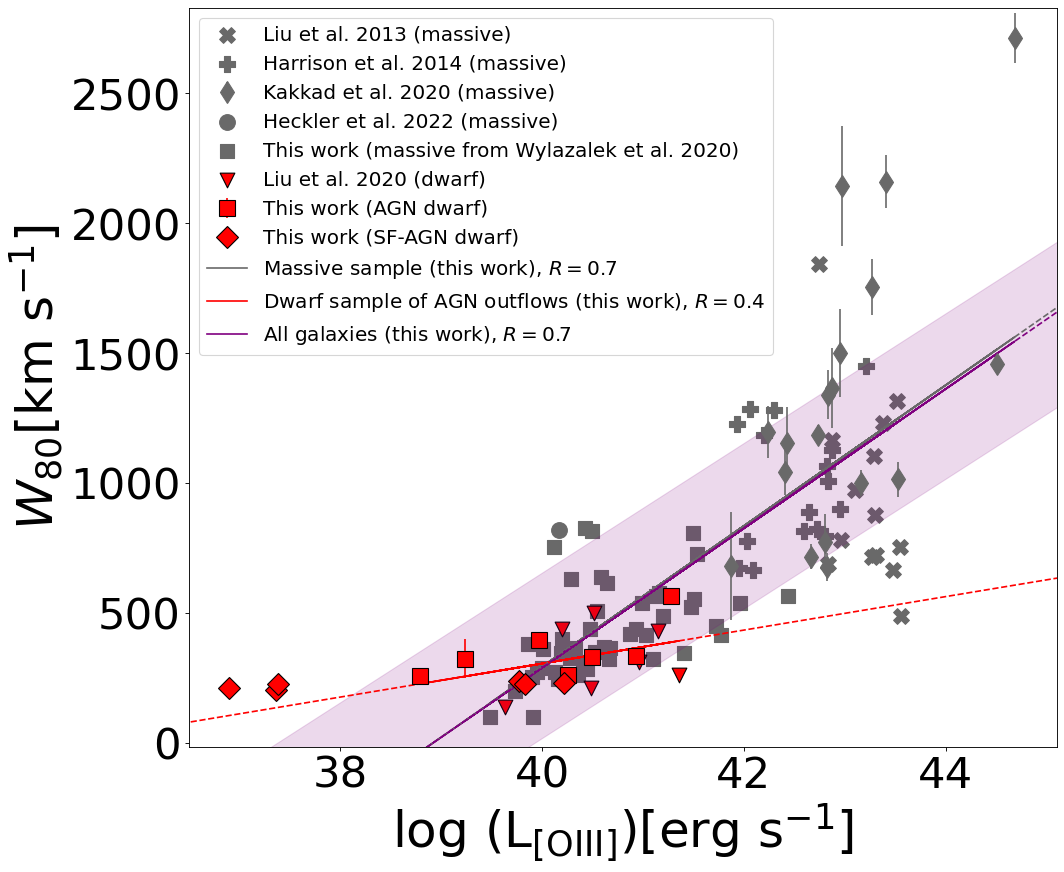} 
\caption{W$_{80}$ velocity vs L$_\mathrm{[OIII]}$ for the MaNGA sample of 13 dwarf galaxies with AGN outflow candidates (red squares for AGN galaxies, red diamonds for SF-AGN galaxies), the AGN dwarf galaxy sample of \protect\cite{liu2020integral} (red triangles), the $z < 2$ type-2 AGN quasars of \protect\cite{harrison2014kiloparsec} (grey pluses), the radio-quiet quasars of \protect\cite{liu2013observations} (grey crosses), the $z \sim 2$ type-1 massive AGN of \protect\cite{kakkad2020super} (grey diamonds),  the massive LLAGN of \protect\cite{heckler2022ifu} (grey circle) and the MaNGA sample of low-intermediate luminosity massive galaxies of \protect\cite{wylezalek2020ionized} (grey squares). The solid lines are regression fits to the massive sample (grey), the dwarf sample of AGN outflows, which includes the nine MaNGA dwarf galaxies with AGN outflows and those from \protect\cite{liu2020integral} (red), and the combination of all the massive galaxies, along with the dwarf sample of AGN outflows (purple) indicated as the shaded area the 1$\sigma$ error of the regression line.}
\label{fig:w80}
\end{figure}

\begin{figure*}
\centering
\includegraphics[width=\textwidth]{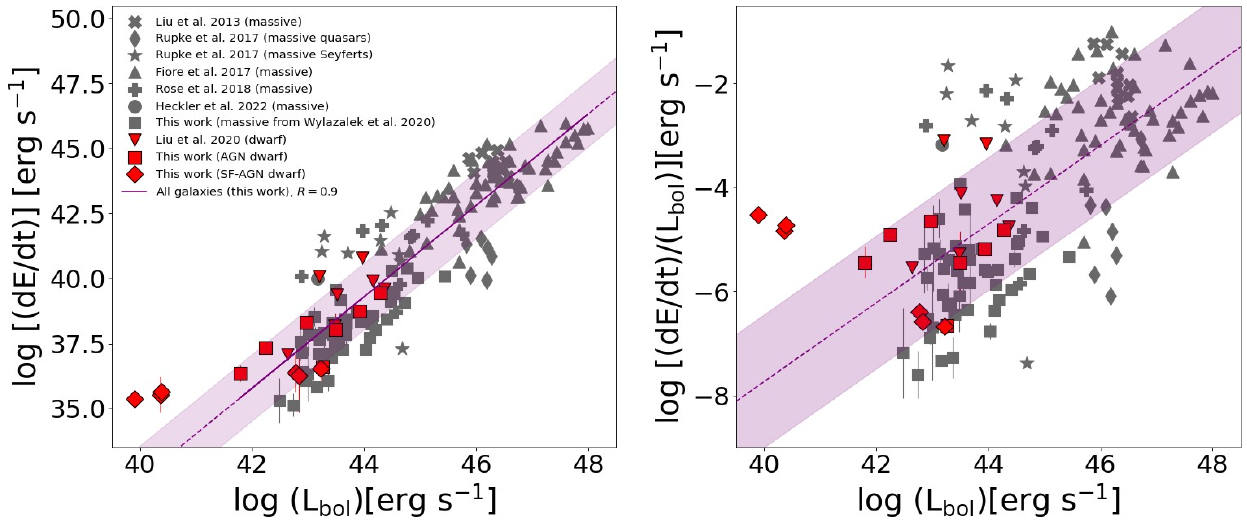} 
\caption{Ionised outflow kinetic energy rate vs AGN bolometric luminosity (left) and ratio of the kinetic energy outflow rate to the AGN bolometric luminosity (right) for the MaNGA sample of dwarf galaxies with AGN outflow candidates (red squares for AGN galaxies, red diamonds for SF-AGN galaxies), the AGN dwarf galaxy sample of \protect\cite{liu2020integral} (red triangles),  radio-quiet quasars of \protect\cite{liu2013observations} (grey crosses),  type 1 quasars of \protect\cite{rupke2017quasar} (grey diamonds), the ULIRGs of \protect\cite{rose2018quantifying} (grey pluses),  MaNGA sample of low-intermediate luminosity massive galaxies of \protect\cite{wylezalek2020ionized} (grey squares), and the massive LLAGN of \protect\cite{heckler2022ifu} (grey circle). The purple solid line is a regression fit considering all the massive galaxies, the nine MaNGA dwarf galaxies with AGN outflows, and those from \protect\cite{liu2020integral}. The shaded area corresponds to the  1$\sigma$ error of the regression line.}
\label{fig:kinrate}
\end{figure*}

Regarding the luminosity, we compared the L$_\mathrm{[OIII]}$ derived from the sum of the AGN and composite spaxels against the total [OIII] luminosity values from the literature; thus, we had to take certain considerations for both dwarf and massive galaxies in this context.\\

For the dwarf galaxies reported in \cite{liu2020integral}, the AGN spaxels contribute at least $\sim 95\%$ of the [OIII] luminosity, which ranges from $\sim 10^{39}$ erg s$^{-1}$ to $\sim 10^{41}$ erg s$^{-1}$, making our results comparable. For the MaNGA sample of nine dwarf galaxies with AGN outflows, we obtain L$_\mathrm{[OIII]}$ that ranges from $\sim$10$^{38}$ erg s$^{-1}$ to $\sim$10$^{41}$ erg s$^{-1}$. These results agree with those in \cite{liu2020integral}. For the four dwarf galaxies with uncertain outflows, three of them present L$_\mathrm{[OIII]}\sim 10^{37}$ erg s$^{-1}$, which is two to four orders of magnitude lower than those from \cite{liu2020integral}, possibly indicating a different origin. \\

In the next section, we compare our results with massive galaxies where the AGN dominates, assuming that the total [OIII] luminosity comes mostly from the AGN.

\begin{table}
\centering
    \caption{AGN luminosity for the MaNGA sample of 13 dwarf galaxies. }

    \begin{tabular}{cccc}
    \hline
    MaNGA & log(L$_\mathrm{OIII}$) & log(L$_\mathrm{bol}$) & log[(dE/dt)/(L$_\mathrm{bol}$)]\\
    plateifu & (erg s$^{-1}$) & (erg s$^{-1}$) &\\
    (1) & (2) & (3) & (4)\\ \hline \hline
        10223-3702 & 40.5  & 43.5 & -5.4\\ 
        10226-1901 & 39.8  & 42.8 & -6.4\\
        11022-12702 & 37.4  & 40.4 & -4.8 \\
        11754-3701 & 40.2  & 43.2 & -6.7  \\         
        11826-12702 & 37.4 & 40.4 & -4.7\\ 
        12487-3704 & 40.3 & 43.3 & -6.6 \\ 
        7992-6102 & 39.8  & 42.8 & -6.6\\   
        8252-1902 & 39.2  & 42.2 & -4.9\\   
        8442-1901 & 40.0  & 43.0 & -4.6\\   
        8655-6101 & 38.8 &  41.8 & -5.4 \\ 
        8657-6104 & 36.9 & 39.9 & -4.5\\ 
        8982-3703 & 41.3 & 44.3 & -4.8\\ 
        9889-1902 & 40.9 & 43.9 & -5.2 \\ \hline

    \end{tabular}
    \tablefoot{ (1) MaNGA plateifu; (2) [OIII] luminosity; (3) AGN bolometric luminosity derived as L$_\mathrm{bol}=1000\times \mathrm{L}_\mathrm{[OIII]}$ (\protect\citealt{moran2014black}); (4) Ratio between the kinetic outflow rate of the outflow and the bolometric luminosity of the AGN}
\label{tab:kineticrate}
\end{table}


\begin{table}
\centering
\caption{Regression lines of the W$_{80}$ vs L$_\mathrm{[OIII]}$ correlation.}
\begin{tabular}{cccc}
    \hline
    Sample & Slope & $R$ & p-value\\ \hline \hline
    Massive sample & 271 $\pm$ 26& 0.73 & $3.9\times10^{-17}$ \\
    \hline
    Dwarf sample & 64 $\pm$ 38 & 0.41 &0.11\\
    \hline
    All galaxies sample& 268 $\pm$ 22 & 0.75 & $1.1\times10^{-21}$\\
    \hline 
\end{tabular}
\label{tab:regression}
\end{table}

\subsection{Comparison with massive galaxies}

In Fig. \ref{fig:w80}, we show the W$_{80}$ velocity versus L$_{\mathrm{[OIII]}}$ for the MaNGA sample of 13 dwarf galaxies with AGN outflow candidates reported here, as well as the dwarf galaxies with AGN outflows studied in \cite{liu2020integral}, the low-to-intermediate luminosity massive galaxies with AGN outflows from \cite{wylezalek2020ionized}, the low-luminosity active galactic nuclei (LLAGNs) with AGN outflows studied in \cite{heckler2022ifu}, the z $\sim$ 0.5 radio-quiet quasars with AGN outflows studied in \cite{liu2013observations}, the z $<$ 0.2 massive type-2 AGN quasars with AGN outflows studied in \cite{harrison2014kiloparsec}, and the z $\sim$ 2 massive type-1 AGN with AGN outflows studied in \cite{kakkad2020super}.  The typical velocity of the AGN outflow candidates in dwarf galaxies appears to be $<500$ km s$^{-1}$, while those of massive galaxies often surpasses the 500 km s$^{-1}$ mark and (in some cases) even  2000 km s$^{-1}$ (e.g. \citealt{kakkad2020super}). However, 29 massive but low-intermediate luminosity galaxies from \cite{wylezalek2020ionized}, whose W$_{80}$ velocities have been recalculated in this paper present velocities of W$_{80}<500$ km s$^{-1}$. We performed a linear regression on the sample of massive galaxies (\citealt{liu2013observations}; \citealt{harrison2014kiloparsec}; \citealt{wylezalek2020ionized}; \citealt{kakkad2020super}), henceforth denoted as the `massive sample'. We also fit a linear regression including just the nine MaNGA dwarf galaxies with AGN outflows in addition to those studied in \cite{liu2020integral} (henceforth `dwarf sample'), along with another one for all the galaxies, massive galaxies, and dwarf galaxies, but excluding the four MaNGA dwarf galaxies with uncertain outflows. Comparing the results of all the regression lines (see Table \ref{tab:regression}), the one that includes all the galaxies and the one representing the massive sample have similar values for both the slope and the correlation coefficient, based on the fact that the regression lines almost overlapped. They have higher correlation coefficient than the one for the dwarf sample. These results could suggest that the AGN outflows in dwarf galaxies can be produced by the same mechanisms as those in massive galaxies, being the dwarf galaxies a scaled-down version of the massive ones (see also \citealt{liu2020integral}). However, the slope of the fit for the dwarf sample is $\sim$ 76$\%$ lower than the one for the all galaxies sample and for the massive sample (see Table \ref{tab:regression}). This may indicate that the W$_{80}$ velocity for the dwarf sample does not follow the same trend as all the galaxies. \\

The kinetic energy outflow rate is a parameter used as a measure of how powerful the outflow is. The values of kinetic energy rate normalised by L$_\mathrm{bol}$ (i.e. outflow efficiency) are shown in Table \ref{tab:kineticrate}. Figure \ref{fig:kinrate} shows, following \cite{liu2020integral}, dE/dt versus L$_\mathrm{bol}$ (left) and (dE/dt)/L$_\mathrm{bol}$  versus L$_\mathrm{bol}$ (right) for the MaNGA sample of 13 dwarf galaxies with AGN outflow candidates. We also refer to \cite{liu2020integral}, \cite{wylezalek2020ionized} (whose kinetic energy rates are derived in this paper), \cite{heckler2022ifu}, and \cite{liu2013observations} in addition to the type 1 quasars with AGN outflows of \cite{rupke2017quasar}, the different AGN at different redshifts (z=0.1-3)  of \citealt{fiore2017agn}, and  ultraluminous infrared galaxies (ULIRGs) with AGN outflows of \citealt{rose2018quantifying}. The bolometric luminosity was calculated from the L$_\mathrm{[OIII]}$ using a bolometric correction factor of 1000 (\citealt{moran2014black}), except for the galaxies in \cite{rupke2017quasar} and \cite{fiore2017agn}, whose L$_\mathrm{{[OIII]}}$ values are not reported and where the bolometric luminosity has been calculated either from L$_\mathrm{bol}=1.15\times\mathrm{L_{IR}}$ (with L$_\mathrm{IR}$ being the infrared luminosity) and from the fitting of the optical-ultraviolet spectral energy distribution (SED). The kinetic energy rate of the 13 MaNGA dwarf galaxies with AGN outflows candidates is four to six orders of magnitude lower than their bolometric luminosity. This suggests that it would be difficult for the outflow to have a significant impact in the galaxy as a source of feedback (\citealt{wang2024rubies}), although there may be additional outflow mass hidden in other phases (\citealt{belli2024star}) such as warm and cold molecular gas phases. This result is consistent with findings related to  the mass-loading factor, where we find that the outflows are unlikely to quench the SF (see Sect. \ref{section:massloading}). Additionally, a linear regression is performed between log(dE/dt) vs log$(\mathrm{L}_{\mathrm{bol}})$, including all the galaxies except for the four MaNGA dwarf galaxies with uncertain outflows. The correlation coefficient is R$=0.9$ indicating that both dwarf and massive galaxies behave similar. In addition to this, a Mann-Whitney U rank test was performed on two independent samples (i.e. massive and dwarf galaxies) returning a p-value > 0.3 in both cases. Consequently, the null hypothesis that the (dE/dt)/Lbol of dwarf and massive galaxies is drawn from the same distribution cannot be rejected. These results may indicate that AGN outflows in dwarf galaxies behave in a similar way as that of massive galaxies, in agreement with the linear correlation when comparing the outflow velocity against the [OIII] luminosity in the `all galaxies sample' of Fig. \ref{fig:w80}. However, the outflow efficiency does not only depend on the AGN luminosity, but also on other factors, such as the  orientation of the outflow with respect to the galaxy disk (\citealt{harrison2024observational}). Thus, a more robust study ought to be performed, which is beyond the scope of the current work.

\section{Conclusions}
\label{section:conclu}

Dwarf galaxies are the most abundant type of galaxies and the building blocks of the more massive ones. Although they were commonly assumed to be regulated by SNe feedback, thousands of AGNs have been found in dwarf galaxies in recent decades. Furthermore, recent studies have shown evidence for the presence of AGN outflows and feedback that may have an impact in the growth of the galaxy. In this paper, the presence of AGN outflows in a sample of 2292 dwarf galaxies, characterised by strong AGN signatures (\citealp{mezcua2024manga}), has been studied. These galaxies are drawn from the MaNGA survey. The stacked spectrum of all those spaxels classified as AGN or composite are fitted using the GELATO python code. For the galaxies where the broad component in the [OIII]$\lambda$5007\AA\ emission line has been fitted, GELATO was run again through all the AGN and composite spaxels to spatially resolve this outflowing component and study the kinematic and energetic properties. We found 13 new  dwarf galaxies with AGN outflow candidates. The main results of this study are:\\

\begin{itemize}
\item{For nine out of the 13 MaNGA AGN dwarf galaxies, there is a misalignment between the gas and the stellar kinematics, which may indicate the presence of potential outflows.}\\

\item{The outflow velocity, W$_{80}$, of the 13 MaNGA AGN dwarf galaxies ranges from 205 to 566 km s$^{-1}$. This result is lower than the typical values of massive galaxies, which often surpass  500 km s$^{-1}$, reaching values above 1000 km s$^{-1}$.}\\

\item{For nine out of the 13 MaNGA AGN dwarf galaxies, the AGN is the most likely mechanism to drive these outflows based on the AGN fraction, the W$_{80}$ velocity, and the L$_{\mathrm{bol}}$ (see Sect. \ref{subsection:origin}). For the remaining four galaxies, both SNe and AGNs are possible candidates for driving these outflows.} \\

\item{The AGN outflow velocity, W$_{80}$, in three out of the {13} MaNGA dwarf galaxies with outflows is higher than the dark matter halo escape velocity. For these galaxies, an enrichment of the circumgalactic medium is expected. }\\

\item{We find a correlation between outflow velocity, W$_{80}$ and L$_\mathrm{[OIII]}$, which  extends all the way from massive to dwarf galaxies with AGN outflows. This may indicate that AGN outflows in dwarf galaxies can be a scaled-down version of massive galaxies. However, the slope of the regression line of the dwarf sample is lower than the one including all galaxies, suggesting that the correlation might not be followed by the AGN outflows in dwarf galaxies.}\\

\item{The kinetic energy outflow rate for the sample of 13 MaNGA dwarf galaxies with AGN outflow candidates ranges from 10$^{35}$ to 10$^{39}$ erg s$^{-1}$. The relation between kinetic energy rate and the bolometric luminosity seems to follow a linear relationship through all the mass regime. However, we find that 3 out of the 13 MaNGA  dwarf galaxies with AGN outflows signatures do not follow this linear relationship, although the origin of the outflow for these galaxies is uncertain. Also, we find that massive and dwarf galaxies are drawn from the same distribution when comparing the outflow efficiency against the bolometric luminosity.}\\

\end{itemize}

In this work, we provide arguments in favour of AGN outflows in dwarf galaxies behaving similarly to those in massive galaxies. However, we find three galaxies that do not follow the linear trends showed in this work, although the driving mechanism that causes the outflow in these three galaxies is not well known. Therefore, the necessity of new detections of AGN outflows in dwarf galaxies is crucial in order to have a better understanding of them, as well as their impact and role in galaxy evolution. In a future work, we plan to study the stellar population properties and examine the effects of SNe in these 13 dwarf galaxies with AGN outflow candidates to further understand the feedback processes and their impact on galaxy evolution.

\begin{acknowledgements}
 The authors thank the anonymous referee for insightful comments. The authors acknowledge the feedback received from Rogemar Riffel and Rogério Riffel from the Universidade Federal de Santa Maria and from the  Universidade Federal do Rio Grande do Sul respectively and from all the colleagues of the AGN and galaxy evolution group of the Institue of Space Sciences of Barcelona. The authors also acknowledge the help provided by Raphael Hviding from the Arizona University and the Max-Planck Institute for Astronomy. VRM acknowledges support from the Spanish Ministry of Science, Innovation and Universities through the project PRE2022-104649. MM and AE acknowledge support from the Spanish Ministry of Science and Innovation through the project PID2021-124243NB-C22. This work was partially supported by the program Unidad de Excelencia María de Maeztu CEX2020-001058-M. HDS acknowledges the financial support from the Spanish Ministry of Science and Innovation and the European Union - NextGenerationEU through the Recovery and Resilience Facility project ICTS MRR-2021-03-CEFCA and financial support provided by the Governments of Spain and Aragón through their general budgets and the Fondo de Inversiones de Teruel. HDS acknowledges financial support by RyC2022-030469-I grant, funded by MCIN/AEI/10.13039/501100011033 and FSE+. AA acknowledges funding from the MICINN (Spain) through the Juan de la Cierva-Formación program under contract FJC2020-046224-I and support by the European Union grant WIDERA ExGal-Twin, GA 101158446. F.M-S. acknowledges support from NASA through ADAP award 80NSSC19K1096. M.S. acknowledges support by the State Research Agency of the Spanish Ministry of Science and Innovation under the grants 'Galaxy Evolution with Artificial Intelligence' (PGC2018-100852-A-I00) and 'BASALT' (PID2021-126838NB-I00) and the Polish National Agency for Academic Exchange (Bekker grant BPN/BEK/2021/1/00298/DEC/1). This work was partially supported by the European Union's Horizon 2020 Research and Innovation program under the Maria Sklodowska-Curie grant agreement (No. 754510).
\end{acknowledgements}

\bibliographystyle{aa} 
\bibliography{Referencias} 

\begin{appendix}
\onecolumn
\section{Effects on S/N threshold selection}
\label{app:A}
\begin{figure*}[h!]
\centering
\includegraphics[width=13cm]{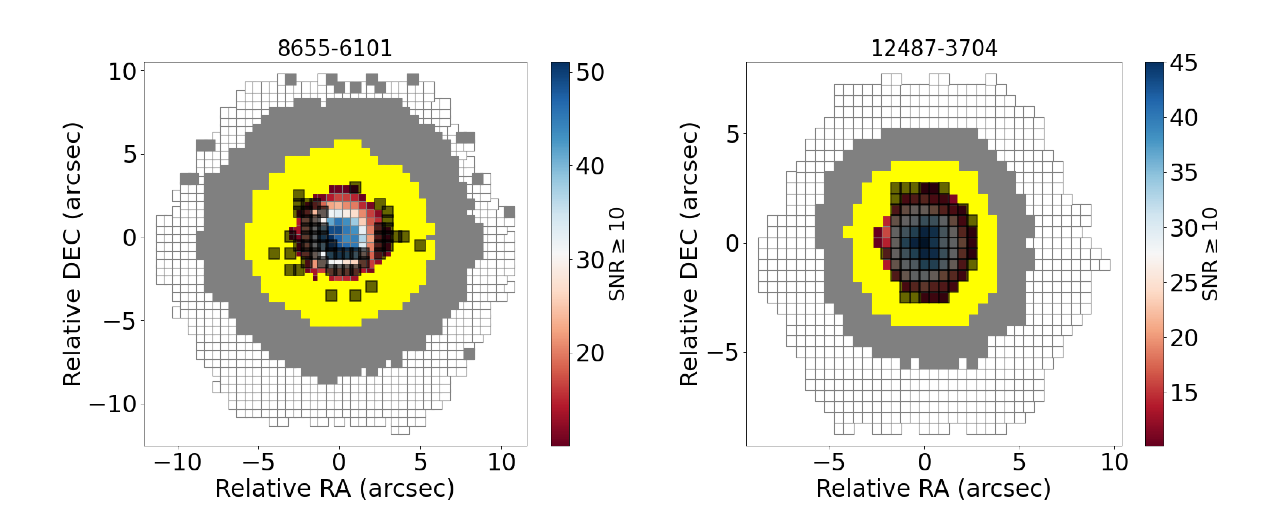} 

\caption{Spatial distribution of the SNR for the same two examples as in Figs. \ref{fig:velmaps} and \ref{fig:velmapsext}. The colourbar is set for those spaxels with S/N$\geq$10. Those spaxels with 3$\leq$S/N$<$10 are shown in yellow to highlight the number of spaxels that would be missed when applying a threshold of S/N$\geq$10 rather than S/N$\geq$3 to ensure the goodness of the fit. The shaded dark spaxels represent the extension of the AGN outflow in each galaxy.}
\label{fig:snr}
\end{figure*}

\cite{mezcua2024manga} performed a spaxel classification only considering spaxels with S/N$\geq$3 in the BPT emission lines (H$\alpha$, H$\beta$, [NII]$\lambda$6583, [SII]$\lambda$6718, [SII]$\lambda$6732, [OIII]$\lambda$5007, and [OI]$\lambda$6300). This is something commonly adopted in BPT studies (e.g. \citealt{reines2013dwarf} for SDSS; \citealt{wylezalek2020ionized} for MaNGA; \citealt{salehirad2022hundreds} for GAMA; and \citealt{johnston2023beyond} for SAMI). However, in our analysis, we consider only spaxels with S/N$\geq 10$ to ensure reliable spectral fittings.  In Fig. \ref{fig:snr}, the spatial distribution of the S/N is shown together with the extension of the outflows of Figs. \ref{fig:velmaps} and \ref{fig:velmapsext} but considering a S/N>3 in the outflow spaxels. No sources with AGN outflows are missed when using S/N$\geq$10 rather than S/N$\geq$3.

\section{Outflow, [OIII]$\lambda$5007, and stellar velocity maps}
\begin{figure*}[h!]
\centering
\includegraphics[width=\textwidth]{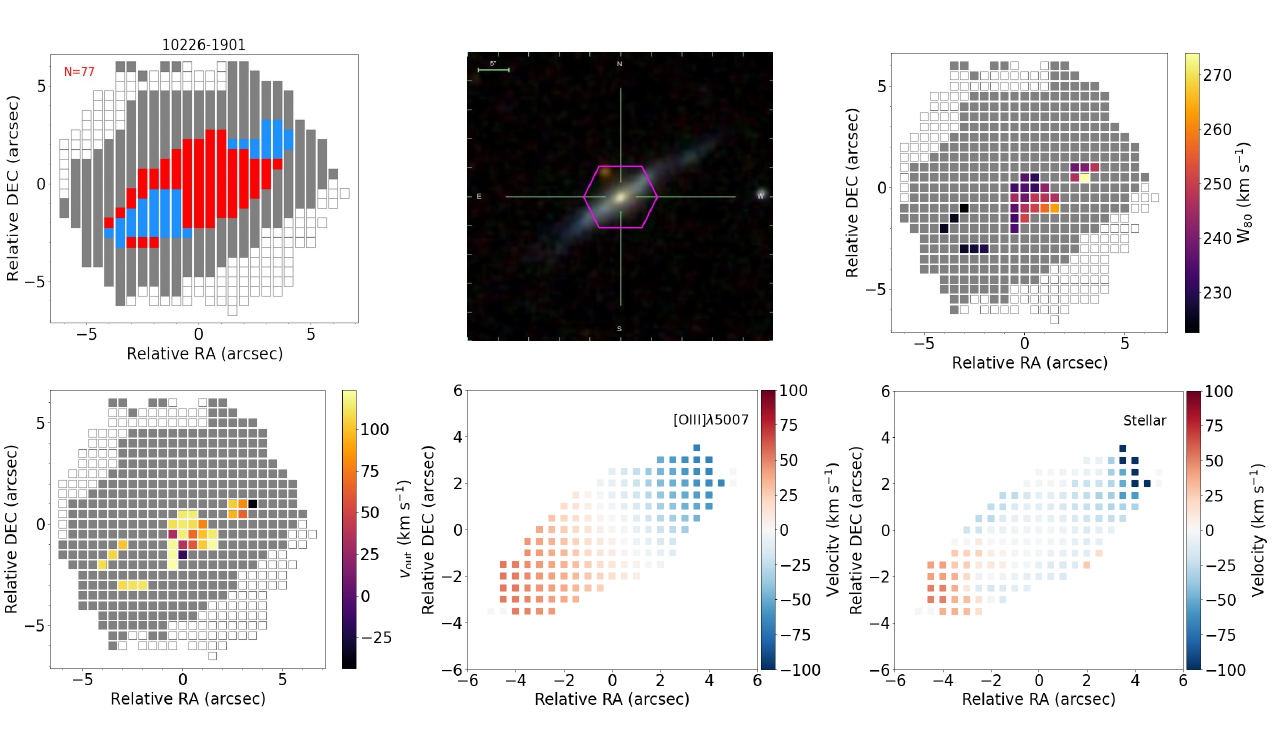} 
\caption{Same as Fig. \ref{fig:velmaps}, but for the dwarf galaxy 10226-1902.}
\label{figure:c1}
\end{figure*}

\begin{figure*}
\centering
\includegraphics[width=\textwidth]{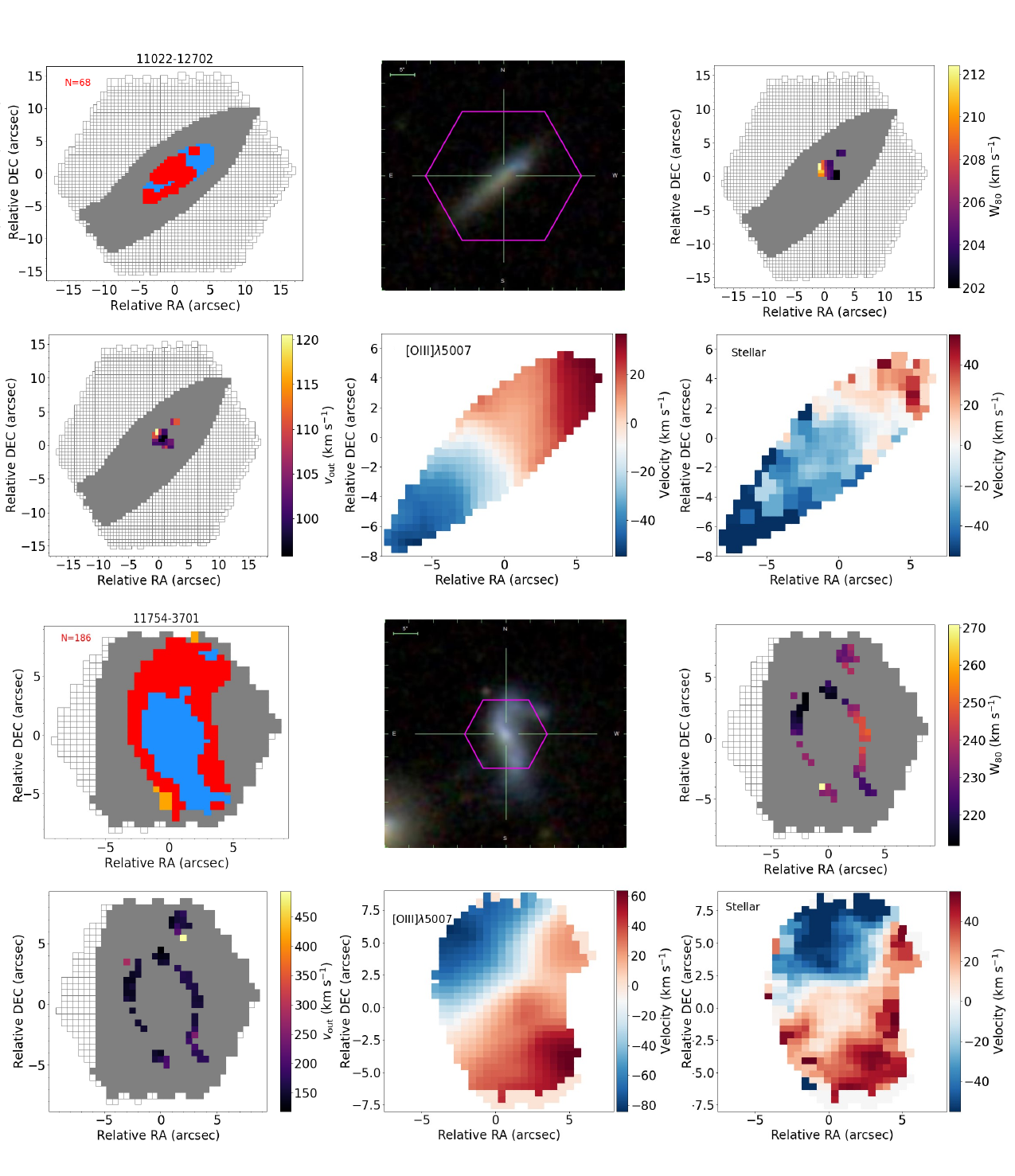} 
\caption{Same as Fig. \ref{fig:velmaps}, but for the dwarf galaxies 11022-12702 and 11754-3701.}
\label{figure:c3}
\end{figure*}

\begin{figure*}
\centering
\includegraphics[width=\textwidth]{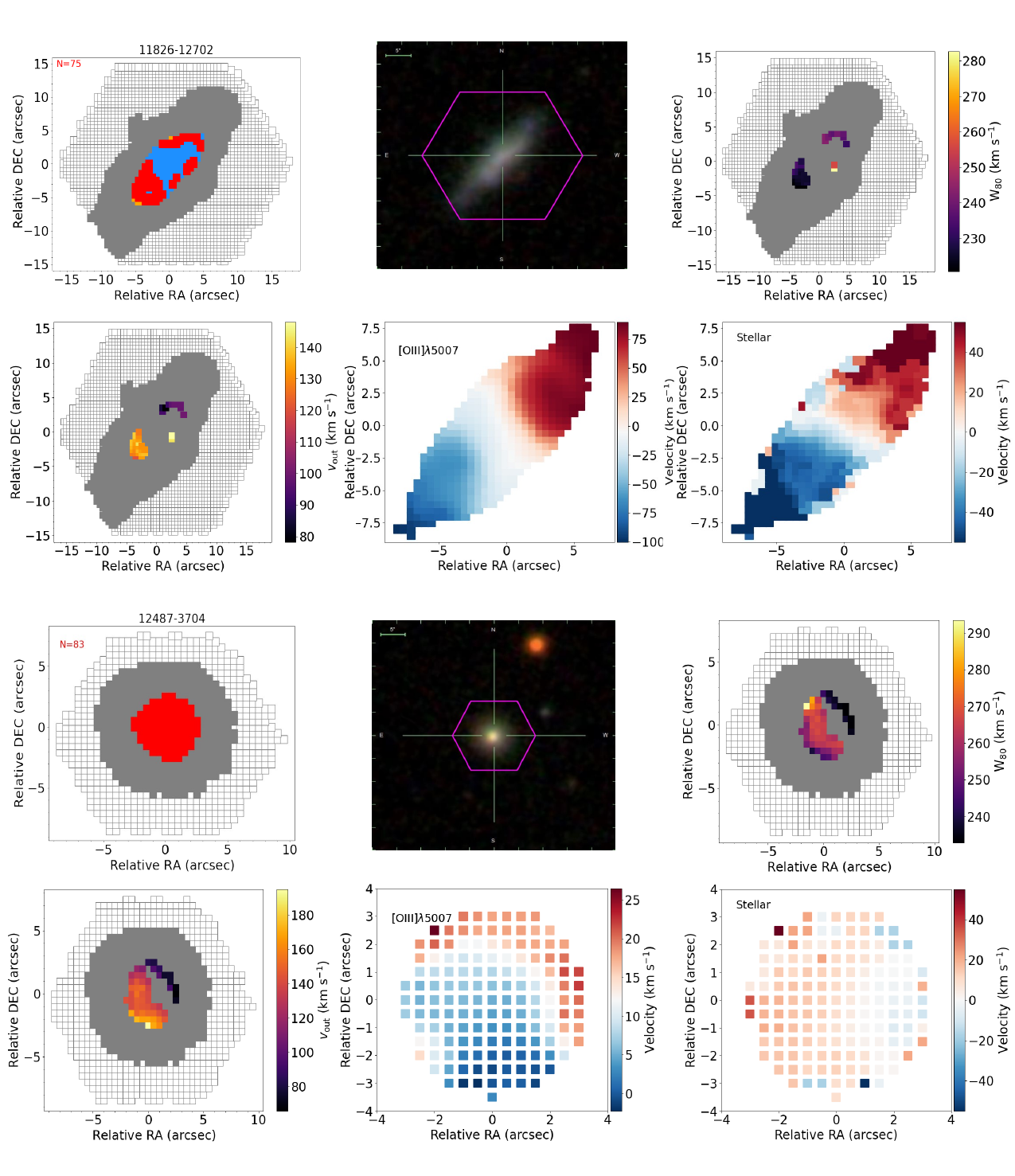} 
\caption{Same as Fig. \ref{fig:velmaps}, but for the dwarf galaxies 11826-12702 and 12487-3704.}
\label{figure:c5}
\end{figure*}

\begin{figure*}
\centering
\includegraphics[width=\textwidth]{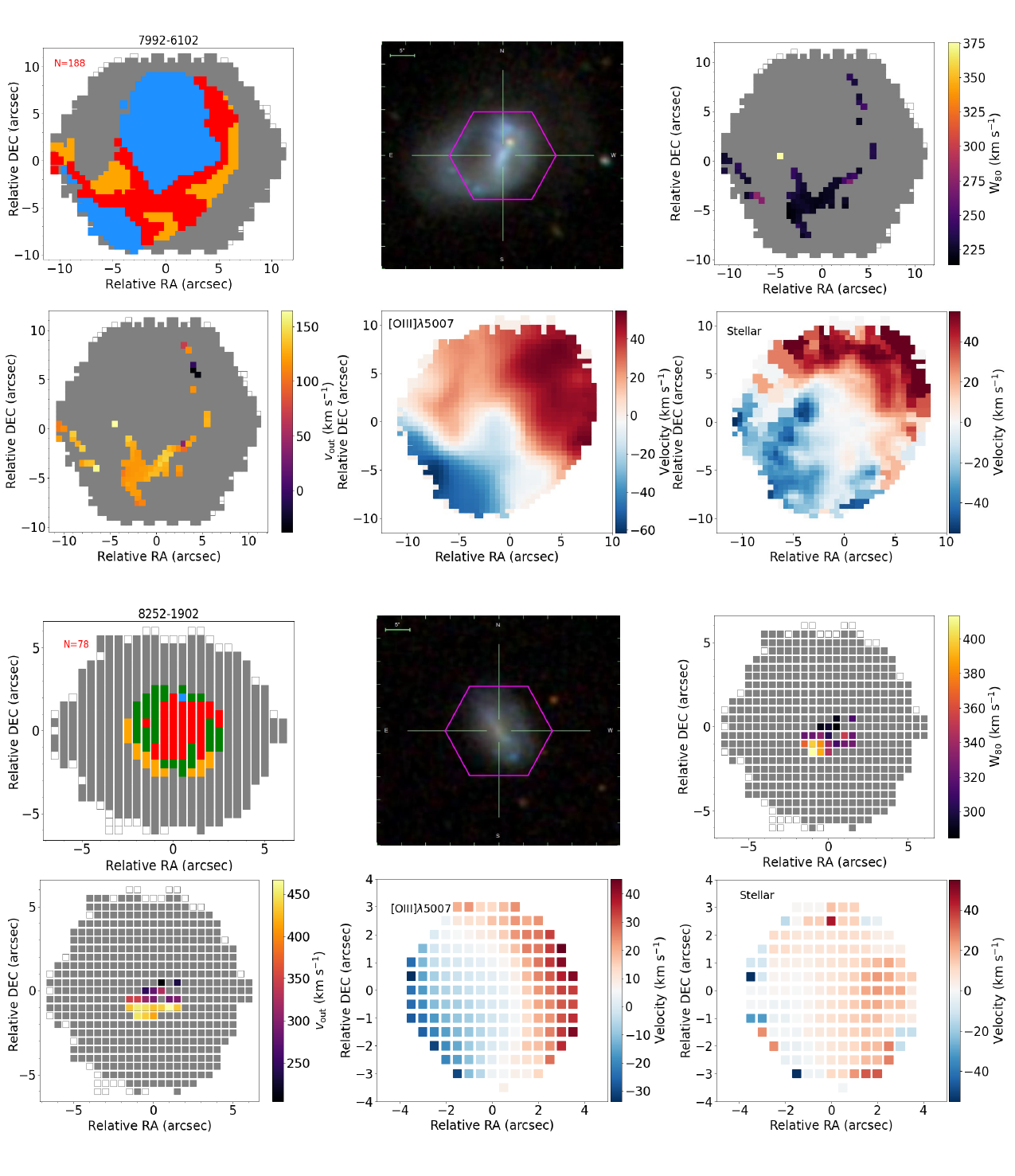} 
\caption{Same as Fig. \ref{fig:velmaps}, but for the dwarf galaxies 7992-6102 and 8252-1902.}
\label{figure:c6}
\end{figure*}

\begin{figure*}
\centering
\includegraphics[width=\textwidth]{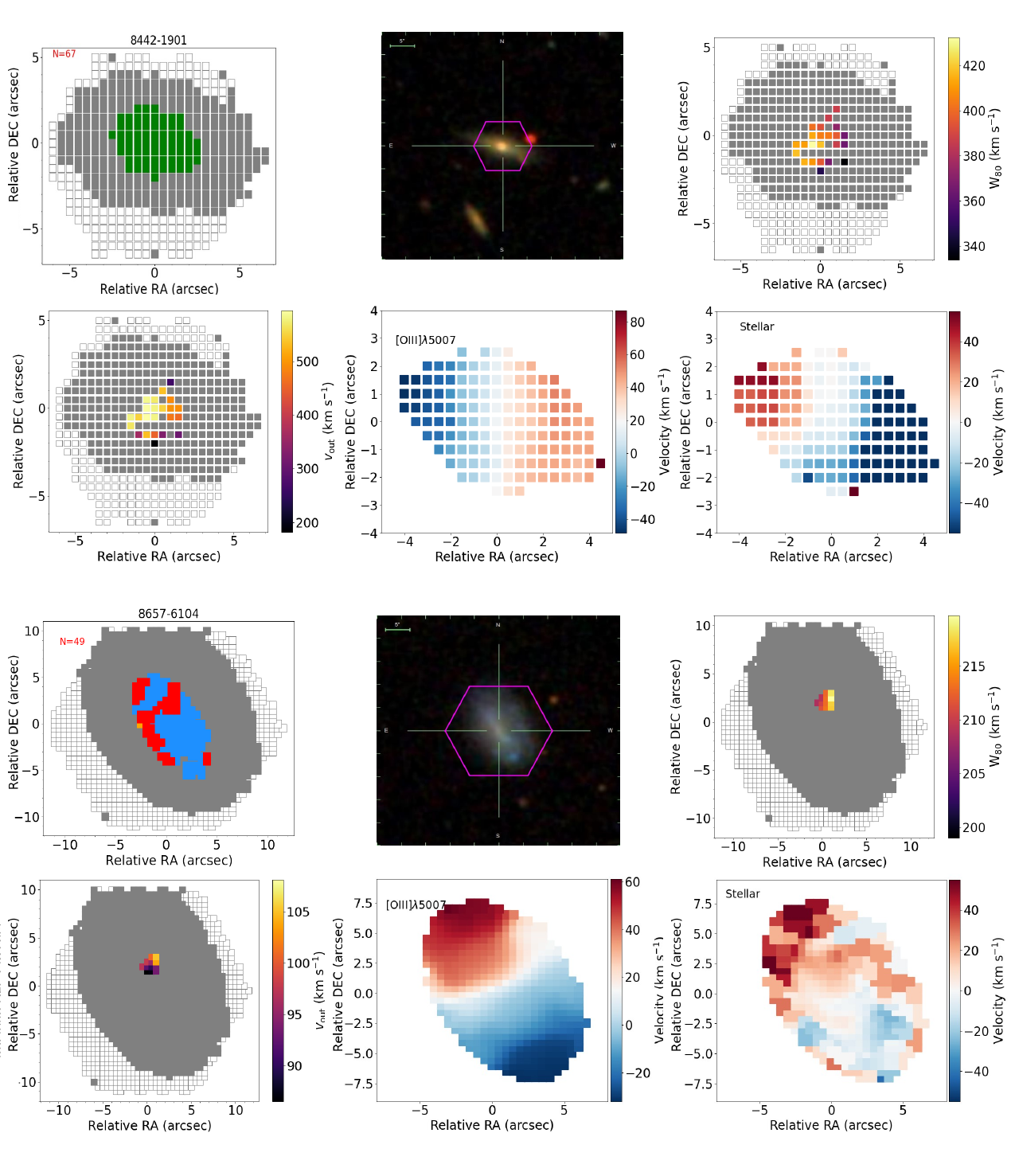} 
\caption{Same as Fig. \ref{fig:velmaps}, but for the dwarf galaxies 8442-1901 and 8657-6104.}
\label{figure:c7}
\end{figure*}

\begin{figure*}
\centering
\includegraphics[width=\textwidth]{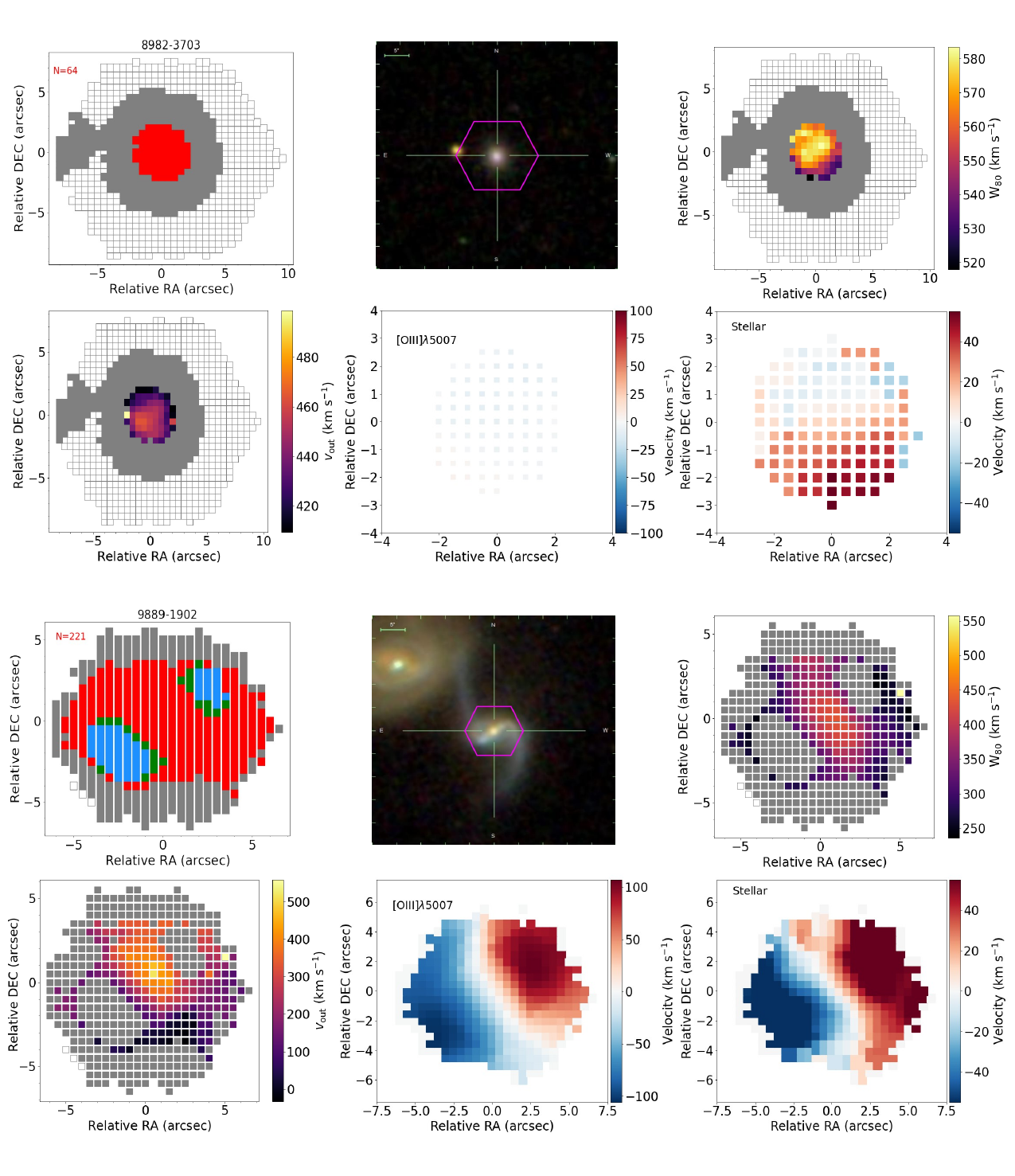} 
\caption{Same as Fig. \ref{fig:velmaps}, but for the dwarf galaxies 8982-3703 and  9889-1902.}
\label{figure:c9}
\end{figure*}

Figures \ref{figure:c1}-\ref{figure:c9} show the outflow, [OIII]$\lambda$5007 and stellar velocity maps of the remaining 11 dwarf galaxies with AGN outflow signatures, as in Figs. \ref{fig:velmaps} and \ref{fig:velmapsext}.

\end{appendix}
\end{document}